\documentclass[lettersize,journal]{IEEEtran}
\usepackage{amsmath,amssymb,amsfonts,amsthm,mathtools}
\usepackage{array}
\usepackage[caption=false,font=normalsize,labelfont=sf,textfont=sf]{subfig}
\usepackage{textcomp}
\usepackage{stfloats}
\usepackage{url}
\usepackage{verbatim}
\usepackage{graphicx}
\usepackage{cite}
\usepackage{xcolor}
\usepackage{blindtext}
\usepackage{lipsum}
\usepackage{subfig}
\usepackage{bigints}
\usepackage{physics}
\usepackage[long]{optidef}
\usepackage{multirow}
\usepackage{steinmetz}

\usepackage{siunitx}

\usepackage{enumitem}
\usepackage{tcolorbox}

\usepackage{algorithm}
\usepackage{algpseudocode}
\usepackage{algpseudocode}

\usepackage{booktabs}
\usepackage{tabularx}
\usepackage{threeparttable}
\usepackage{pifont}
\newcommand{\cmark}{\ding{51}}
\newcommand{\xmark}{\ding{55}}
\definecolor{addblue}{RGB}{0,90,200}
\definecolor{delgray}{RGB}{140,140,140}
\definecolor{notered}{RGB}{200,30,30}
\newcommand{\rnote}[1]{}                                  

\begin{document}

\title{EBGT: Epistemology-aided Bayesian Game Theory for Uplink Power Control in Stochastically Distributed IoT Tiers}

\author{Nirmal~D.~Wickramasinghe,~\IEEEmembership{Student~Member,~IEEE,}
Indrakshi~Dey,~\IEEEmembership{Senior Member,~IEEE,}
Dirk~Pesch,~\IEEEmembership{Senior Member,~IEEE,} 
and John~Dooley,~\IEEEmembership{Member,~IEEE}

\vspace{-11mm}

\thanks{N.~D.~Wickramasinghe is with the Department of Electronic Engineering, Maynooth University, Ireland. (Email: nirmal.wickramasinghe.2023@mumail.ie)}
\thanks{I.~Dey is with Department of Computing and Mathematics, South East Technological University, Waterford, Ireland. (Email: indrakshi.dey@setu.ie)}
\thanks{D.~Pesch is with the School of Computer Science and Information Technology, University College Cork, Ireland. (Email: d.pesch@cs.ucc.ie)}
\thanks{J.~Dooley is with the Department of Electronic Engineering, Maynooth University, Ireland. (Email: john.dooley@mu.ie)}
}



\maketitle
\begin{abstract}
Uplink power control in dense, heterogeneous Internet-of-Things (IoT) tiers is fundamentally limited by incomplete channel-state information (CSI) and mutual interference, while low size, weight, and power (SWaP) devices cannot afford the feedback and computation of conventional distributed schemes. This paper proposes EBGT, an epistemology-aided Bayesian game-theoretic framework for decentralized uplink power minimization in stochastically distributed IoT networks. Interfering users are modeled as spatially random through a Poisson point process (PPP), and each device reasons about its rivals through a two-layer belief hierarchy of \emph{inter-epistemic} beliefs about opponents and \emph{intra-epistemic} self-assessment, so that the transmit-power equilibrium is reached without repeated inter-node feedback. We derive a closed-form coverage-probability payoff via stochastic geometry and quantify belief convergence toward equilibrium using the Jensen--Shannon divergence (JSD) of the resulting power distributions. Monte-Carlo simulations validate the analytical coverage expressions and show that EBGT sustains the target coverage probability while reducing transmit power relative to fixed power control (FPC) and stochastic non-cooperative power control (SNCPC) baselines, particularly under stringent SINR and high-density regimes.
\end{abstract}

\vspace{-5mm}
\section{Introduction}
\IEEEPARstart{N}{ovel} IoT technology is growing rapidly on top of advanced embedded systems in parallel to big data analysis in the environment introduced through diverse generative models, \cite{IoT_LLM}. And, there is a remarkable surge of IoT entities in the network, such as Ericsson's latest mobility report \emph{(June-2026)}, \cite{EricssonMobilityReport2026} forecasts that the total number of cellular IoT connections was around 4.5 billion at the end of 2025 and is expected to approach 8 billion by the end of 2031. 
Moreover, multi-task IoT packagings introduce dynamic IoT widgets to accelerate learning models and demand game-changing uplink resource allocation technologies.

\subsection{IoT Uplink Resource Allocation and Challenges}
IoT devices are energized by a primary battery and consume power to sense physical parameters surrounding and processing actuated data. Essentially, IoT nodes spend the vast majority of the energy budget on conveying data packets among tranceivers, \cite{IoT_pow_cons_01, IoT_pow_cons_02}. Specifically, IoT firings drain a higher percentage of the battery than the downlink against distance-related path-loss attenuation and hardware impairments, \cite{IoT_uplink_power_thant_downlink}. The uplink constitutes the primary energy bottleneck in wireless networks due to the limited battery capacity of user equipment, whereas downlink transmissions are supported by IoT gateways or base stations with continuous and substantially less constrained power supplies. The ideal task of power regulation is a convenient and signal-to-noise ratio (SNR) related function in conventional IoT platforms, assuming orthogonal space-time-frequency resource blocks and no collisions. 

Nevertheless, effective resource management remains a significant challenge in applied deployment settings that saturate the available spectrum with a huge number of devices in the network. Hence, the latest IoT subscriptions are compacted in the spectrum and attain the fair band gap among the carrier frequency slots, \cite{Low_band_gap_vs_intereference}. On the other hand, service providers are proposing use case-specific resource scheduling and queueing protocols that would increase the frequency reuse factor \cite{freq_reuse_01, freq_reuse_02, Queuing_Theory_RA}. And, device heterogeneity would raise uncertainty in critical stages, and therefore, that would imbalance standard resource arrangements and generalized limited constraints, \cite{IoT_heterogeiny_challenge}. For instance, low power wide area network (LPWAN) technologies, i.e, low range wide area network (LoRaWAN) \cite{LoRaWAN}, narrow band (NB-IoT) \cite{NB-IoT}, and novel Sigfox IoT \cite{Sigfox} communication protocols would utilize identical industrial, scientific and medical (ISM) purpose bands, such as in Europe, the 863 MHz to 870 MHz range is designated for license-free communications for short-range devices (SRDs), \cite{EU_unlicensed_freq_band}. Moreover, the mobility of IoT users changes the network topology and introduces additional dynamic channel impairments on top of channel fading uncertainties caused by surrounding obstacles. Hence, these numerous burdens could result in unexpected overlaps and time-frequency jitters toward consecutive resource blocks, creating mutual interference at the IoT gateway among IoT user uplinks. Now the uplink resource management problem is no longer a zero-force interference approach, and the goal is to maintain a sufficient data rate under energy constraints while minimizing interference leakages. 

\subsection{IoT Interactions and Game Theory}
The resource allocation literature is comprehensive, drawing on classical analytical approaches, iterative trials, and sequential algorithms. Nonetheless, most of the literature that could enhance the local performance metrics lacks the potential to capture complex interactions among users in the wireless network, leading to degradation of the entire performance of the system. An example of IoT uplinks are motivated to increase the transmit power gain to suppress the mutual interference noise and accomplish channel throughput targets for better quality of service (QoS) in communication, \cite{GT_based_RA, ericsson_rl_powercontrol}. The excessive power on top of the received signal strength indication (RSSI) at the IoT gateway would strengthen the cumulation of interference leakage, which could punish the neighbour uplinks with weak channel gains and vice versa. Hence, it is essential to introduce interoperable service protocols with the capability of rational decision-making that would impact for the entire cyber-physical system (CPS), \cite{rational_interoperable_IoT}. Game theory serves as a market-oriented mechanism to model the resource management problem, analyzing the strategic behavior of users to take reasonable judgments. There are different types of game models regarding the \emph{information awareness}: complete and incomplete, \emph{entity cooperation}: cooperative and non-cooperative games that would depend on the edge computing metrics, \cite{Game_theory_auction_edge}. The performance accuracy, complexity, and feasibility of the typical framework rely on the strategic nature of the game model, and non-cooperative behaviour with problem uncertainty generalized the most challenging yet viable use cases in the wireless edge network.

Incompleteness allows rational players to follow a given common prior distribution to model user interactions and then make conditional strategies to maximize the average utility space of the system. This can be identified as Bayesian game theory (BGT), and there are numerous specific literature in wireless networks, such as resource management \cite{BGT_RA_01, BGT_RA_02, BGT_RA_03}, security \cite{BGT_security_01, BGT_security_02, BGT_security_03, BGT_security_04}. The problem of uplink power minimization and preserving the channel throughput threshold to enhance the network coverage aligns well with the principles of BGT.  Here, coherent IoT nodes select the best minimal power strategy while maximizing the average utility throughput of the network, against interfering neighbours, sustaining unknown channel state information, \cite{BGT_uplink, my_GLOBECOM_BGT}. Despite IoT entities being able to reach global uplink optimality of the network service selection following BGT, it is expensive to compute expected utility tables exponentially along the size of the network, action space, and amount of uncertainty prior to exploring the equilibrium. This motivates the design of novel service allocation protocols based on rational strategies that are compatible with the processing constraints of network entities.

\subsection{Operational Paradigms and Edge Intelligence (EI)}

\begin{table}[t]
\centering
\caption{Comparison of IoT-uplink resource-allocation frameworks}
\label{Tabl: Uplink_RA_literature}
\begin{threeparttable}
\begin{tabularx}{\columnwidth}{@{}>{\raggedright\arraybackslash}X c c c c c c c@{}}
\toprule
\textbf{Paper} & \textbf{Method} & \textbf{CSI} & \textbf{Mob.} & \textbf{Intf.} & \textbf{Cgs.} & \textbf{FB/Cx}\\
\midrule
\cite{my_GLOBECOM_BGT}  &  BGT-classical  &  I   & \xmark  &  \cmark  &  Sim  &  High \\
\cite{Stackelberg_game_RA}  &  Stackelberg Game  & I   & \xmark  &  \xmark  &  Sim   &  FB-Med \\
\cite{Auction_Theory_RA_myone}  &  Auction theory  & I  & \cmark   & \cmark  &  \cmark  &  FB-Med\\
\cite{Eff_RA_MEC}  &  PSO \& MINLP  &  C  & \xmark   & \cmark  &  Sim  &  Med\\
\cite{RSMA_URLLC}  &  Statistical \& TSSO  & I   & \xmark   & \cmark  &  Sim  &  Low\\
\cite{MFG_grant_free} & MFG, grant-free  & I  & \xmark   & \cmark  &  Sim  &  FB-Med\\
\cite{MFG_traffic_aware}  &  MFG, traffic-aware  & I  & \xmark  &  \cmark  &  \cmark   &  Med \\
\cite{Stochastic_Markov_game}  &  Stoch. game (D2D)  & C  & \cmark  &  \cmark  & \cmark  &  Lin-High\\
\cite{DRL_NOMA}  &  DRL / NOMA  & C  & \xmark   & \cmark  &  Sim  &  Tr-High\\
Our &  \textbf{EBGT-Stoch.}  & I  & \cmark   & \cmark  &  \cmark  &  Low \\
\bottomrule
\end{tabularx}
\begin{tablenotes}
\scriptsize
\item[] 
\textbf{CSI}: BGT=Bayesian-Game-Theory, PSO=Particle swarm optimization, MINLP=Mixed-integer-non-linear-programming, TSSO=Three-step sequential optimization, MFG=Mean-field-game, Stoch.=Stochastic, D2D=Device-to-device, DRL=Deep reinforcement learning, NOMA=Non-orthogonal-multiple-access.
\textbf{CSI}=Channel-state-information: C=complete, I=incomplete/imperfect.
\textbf{Mob.}=User mobility.
\textbf{Intf.}=Presence of SINR interference.
\textbf{Cgs.}=Guarantee for convergence: Sim=Simulation, \cmark=Theoretical/Analytical.
\textbf{FB/Cx}=Feedback \& complexity: Lin-High=Linear with number of players, Tr-High=Training heavy
\end{tablenotes}
\end{threeparttable}
\end{table}
A variety of core paradigms have been developed for strategic resource allocation in IoT uplink networks, including user-centric, user-distributed, and pre-trained model-based approaches, among others, to address the requirements of diverse use cases. The user-centric frameworks make internal decisions about transmission despite the user dynamics, shaded metrics, and lack of processing gain would result in inefficient conclusions and considerable delays. Typically, IoT devices are the simplest platform in physical layer in the wireless network and recalling low size, weight, and power (SWaP) entities with a lack of processing capabilities. Hence, user-distributed IoT resource sharing architectures are introduced in the literature, such as offloading heavy computation task to cloud layers \emph{cloud-layer}, \cite{IoT_cloud_RA} and third party mobile edge computing (MEC) platforms \emph{Fog-layer}, \cite{IoT_fog_RA_01, IoT_fog_RA_02}. The legacy of game theory discuss, distributed decision-making platforms named Stackelberg game \emph{leader-followers} \cite{Stackelberg_game_RA} similar to the operation in auction models dealing among auctioneer and bidders, \cite{Auction_Theory_RA_myone}. Additionally, instantaneous cluster-based decision makers enhance the entire group utility \emph{(Shapley value)} while sharing the ultimate outcome with respect to the percentage of coalition from each sub-space \cite{Coalitional_game_RA}. The user-distributed architectures are powerful to make actions cooperatively with the benefits of parameter awareness, ability to regulate complexity for reduced-capability (RedCap) that is compatible with glimpse of 5G new radio (NR) systems \cite{RedCap_Ericsson_2023}. However, these advantages are accompanied by several inherent challenges, including additional idle time slots required for acknowledgment and feedback exchange, synchronization overhead, and the potential for monopolistic decision-making that may lead to unfair resource allocation. In contrast, alternative approaches employ machine learning (ML) and artificial neural networks (ANNs) in advanced IoT processing frameworks such as \emph{TinyML} and \emph{Edge-AI} \cite{tinyML_IoT, EdgeAI_IoT_RA_01, EdgeAI_IoT_RA_02}. However, node-level performance improvements are limited by scalability and deployment cost constraints, as well as by a fundamental trade-off between the size of the pre-trained dataset and inference accuracy, which may become fragile under unexpected operating conditions. The table \ref{Tabl: Uplink_RA_literature} compares diverse resource allocation approaches for IoT uplink. 

According to the discussion of operational paradigms for network service algorithms, it is challenging to propose edge-biased and computationally distributed frameworks without relying on repeated feedback to guide navigation toward optimality. In response to these limitations, decentralized protocols incorporate virtual collaboration among network entities by exploiting cognitive and behavioral interaction patterns within the network, a concept commonly referred to as socialized learning (SL), \cite{SL_EI}. Subsequently, end nodes leverage social principles that strengthen the collaborative decision-making capabilities of edge agents known as edge intelligence (EI) beyond conventional data-driven repetition and pre-trained knowledge structures. The fusion of EI and SL enables the IoT edge to process mutual inference efficiently, enhancing the system intelligence. In addition, emerging signaling schemes are more productive as per split learning among IoT entities through statistically shared experiences, whereas improving the adaptability in complex and dynamic environments by enabling context-aware interactions, thereby enhancing end-user experience. Consequently, an intuitive belief-guided BGT-based resource allocation framework for edge computing, which captures and balances the mutual interactions among split-learning modules under a shared prior assumption.

\subsection{Epistemology for BGT (EBGT)}
Epistemology philosophically justifies the theory of knowledge by investigating the relationship between belief, truth, and the degree of confidence required for a claim to be regarded as knowledge. The SL explicitly encourages IoT users to make mutual beliefs against opponents in the intuitive reflection of \emph{what players believe about... what others believe about rationalities of players}, \cite{epistemology_phrase}. The legacy of BGT builds expected utility tables sequentially for each player against the conditional unknown opponent state space and strategy profile prior to executing the task of expected utility maximization. This presents recurrent utility spaces for unnecessary exploration and demands extensive processing power to conclude the equilibrium state. Despite the immense computations, exploited by game incompleteness with continuous common priors, players would tailor beliefs that strengthen the central idea behind Bayesian epistemology, which is identified as degrees of belief (DoB). The DoB profile adheres to the standard probabilism of non-negativity, sum-to-one, and additivity for the possibilities in the belief proposition, \cite{epistemology_Bayesian_Stanford_01}. In the EBGT framework, the EI makes arguments among end users, splitting the SL to make beliefs about the opponent layer by each desired user \emph{(inter-epistemic beliefs layer)}, thereby self-assessments \emph{(intra-epistemic beliefs layer)}. Subsequently, the mutual belief layers are stacked to build the statistical belief hierarchy, extracting user interactions, and Section \ref{Sec: Epistemology_for_Bayesian_Games} provides an explicit discussion through logical reasoning and graphical conceptualization. In the context of a belief hierarchy, hypothesis are \emph{mutually exclusive}: avoid two possibilities holding together and \emph{jointly exhaustive}: holding at least one of the possibilities. To navigate the inter-intra epistemic belief hierarchy, the hypothesis with the highest DoB is inductively inferred to the evidence stage in aid of the conditionalization principle, \cite{epistemology_Bayesian_Stanford_01}. The statistical epistemic space is captured by the incompleteness of IoT spatial dynamics through stochastic geometry, providing mathematical tractability.     

\subsection{Stochastic IoT Networks}
To capture border range of physical metrics and coverage enhancement, IoT edge is dynamic, including cellular IoT, smart wearables, Internet of vehicles (IoVs), and non-terrestrial network (NTN) applications, \cite{IoT_random_movements}. Stochastic geometry (SG) introduces mathematically amicable properties to model random spatial patterns of movable users through point processes (PP). Each desired IoT device establishes epistemic beliefs on the kernel of a homogeneous Poisson point process (PPP), to characterize the stochastic IoT dynamics in the presence of opponent interference. PPP is a robust and fair approach to implement uncertainty behaviours of IoT end users that would satisfy the empirical network coverage performance with analytical derivations, \cite{PPP_match_practical, Primer_Dhillon}.

\subsection{Contributions}
In this paper, we address critical challenges in the resource allocation problem for distributed IoT networks. This is an extended network service management protocol that was initiated from our previous work explains in \cite{my_EBGT}, and the first-ever stochastic epistemology-inspired lightweight BGT resource allocation architecture for user-interactive IoT uplink power controls. The essential contributions of this paper are as follows.
\begin{itemize}
    \item We examine a heterogeneous IoT network where edge users are expecting diverse channel throughput bounds for the simultaneous uplinks compared to conveying packet sizes. Additionally, we account for the fact that the present resource allocation problem suffers from exponential complexity with the network size, the degree of CSI incompleteness, and the strategy profile and neighboring interfering users, warning to maintain a proper manner of transmit power regulation that would preserve the achieved network coverage. 
    \item To address challenges encountered, we propose a novel efficient resource management protocol that would be suitable for low-power and memory devices operating under real-time dynamics. The desired framework leverages Bayesian epistemology to infer intelligence-aided decision-making about competitive IoT interferences over data-driven sequential learning. Although the distributed EBGT architecture makes cooperative decisions in the IoT edge, the iterative convergence is decoupled from external intermediate state updates and seeks the equilibrium blindly in aid of DoB from hypothesis to evidence realizations. 
    \item For a clearer understanding of the EBGT evolution, we first cast the optimization problem as an imperfect game with complete knowledge but unaware opponent strategy profile, then introduce a statistical cognitive hierarchy. Here the rational IoT decision maker tailors inter- and intra-epistemic belief layers against rivals and self pre-states, respectively, and converges toward the global optimum, exploring expected coverage utilities. We discuss the inter-intra-epistemic rationality of the stratified belief system and measure the development of belief correlation toward Nash equilibrium via a metric deviation heatmap of Jensen-Shannon Divergence (JSD).
    \item We consider a severe chance of neighbor interference for the desired user uplink owing to a higher frequency reuse factor in the vicinity for spectrum saving and scaling subscription offerings. The IoT end users are scattered in a PPP with random spatial patterns under SG to align with the realistic network dynamics in the resource allocation problem. We evaluate the internal rationality of the EBGT framework and verify whether the derived inter- and intra-epistemic stochastic analytics \emph{(\figurename~\ref{Fig: P_coverage_vs_P_k_inter_epistemic} and \figurename~\ref{Fig: P_coverage_vs_P_i_intra_epistemic})} are well-suited to practical Monte-Carlo simulations.  
    \item Leading to the overview of the proposed algorithm, we analyze the performance of the IoT uplink power minimization with respect to the desired tagged user distance for the network. Moreover, we illustrates the average transmit power gains against three pillars of network metrics such as coverage probability bounds, SINR threshold, and the PPP mean user density. Thereafter, each desire IoT player is firing to the connected gateway employing the power strategy profiles that were obtained from the proposed EBGT protocol and evaluating the achievable network coverage probability. Ultimately, we show that the functionality of our desired EBGT protocol shows robust overall performance compared to two specific IoT uplink power control benchmarks, specifically fixed power control (FPC) and stochastic non-cooperative power control (SNCPC).
    \end{itemize}

\section{System Model and Problem Statement}
\begin{figure*}[!t]
\centering
\includegraphics[width=0.8\linewidth]{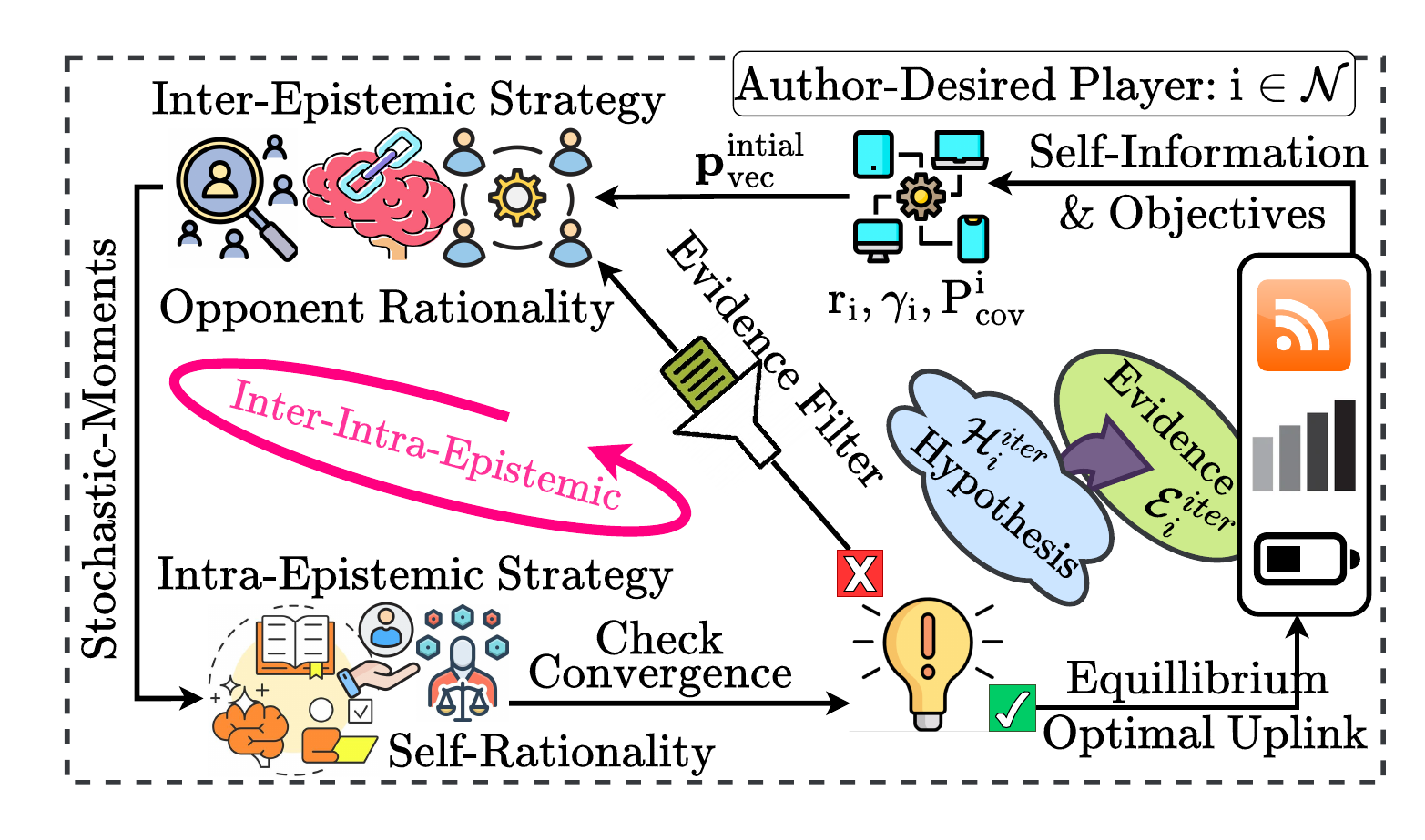}
\centering
\caption{Representation of the conceptual Epistemic Bayesian Game Theory (EBGT) model of the desired IoT node scattered in the stochastic PPP network}
\label{Fig: EBGT_System_model}
\end{figure*}
The dynamic IoT devices are spread across the distributed network and connected to the desired IoT gateways which lie on a marked PPP $\mathrm{\Phi}$, with density $\mathrm{\lambda}$, which is built upon multiple PPP tiers. During uplink, the typical IoT node $\mathrm{i \in \mathcal{N}}$ transmits sensed and actuated zero-mean unity power data symbol vectors $\mathrm{\mathbf{s}_{i} \in \mathbf{S} \subset \mathbb{C}^{2}}$ randomly where, energized by $\mathrm{p_{i} \in \mathbf{p}}$ power strength. Then the received signal $\mathrm{\mathbf{y}_{i}}$ at the $\mathrm{i^{th}}$ typical IoT gateway located at a distance $\mathrm{r_{i}}$, can be mathematically formulated as,
\begin{align}\label{Eq: rx_sig}
   \mathrm{ \mathbf{y}_{i} = \sqrt{p_{i}} \mathbf{s}_{i} \mathbf{g}_{i} r_{i}^{-\alpha} + \sum_{j \in \mathcal{N} \setminus i} \sum_{\mathbf{x} \in \mathbf{\Phi}_{\mathcal{I}_j}} \sqrt{p_{j}}  \mathbf{s}_{j} \mathbf{g_{\mathbf{x}}} \left\| \mathbf{x} - \mathbf{o}_{i} \right\|^{-\alpha} + \mathbf{n}_{i}}
\end{align}
where $\mathrm{\mathbf{g} \in \mathbf{G} \subset \mathbb{C}^{2}}$ is the corresponding complex 2D iid Gaussian fading vector. The ubiquitous frequency reuse will enhance the number of subscriptions in the IoT network; however, it would introduce frequent collisions at the typical IoT gateway. These  heterogeneous IoT interference clusters follow individual PPP distributions $\mathrm{\Phi_{j}}$, with density $\mathrm{\lambda_{j}}$ and $\mathrm{\lambda = \sum_{j \in \mathcal{N}\setminus i} \lambda_{j}}$. Each IoT interferer is located at $\mathrm{\mathbf{x} \in \mathbf{X} \subset \mathbb{C}^{2}}$ positions in respective clusters and collided with the typical IoT gateway $\mathrm{i}$, originated  at $\mathrm{\mathbf{o_{i}} \in \mathbf{O} \subset \mathbb{C}^{2}}$. Uplink signals are attenuated along the standard power-law path loss exponent and is denoted by $\mathrm{\alpha}$ and distorted by the fundamental zero-mean additive white Gaussian noise floor $\mathrm{\mathbf{n}\sim \mathcal{N}(0,\sigma_{n}^{2})}$.  

Although the desired user $\mathrm{i}$ is aware of self-CSI $\mathrm{\mathbf{g}_{i}}$ and distance from the tagged IoT gateway $\mathrm{r_{i}}$, it is challenging to acquire deterministic information on instantaneous locations $\mathbf{x}$ of IoT interferences and CSI-themselves $\mathrm{{\mathbf{g_x}}}$. Therefore, we can pose the challenging uplink power minimization problem while satisfying the given SINR throughput threshold $\mathrm{\gamma^{th}}$, \eqref{Eq: Constraint_SINR}, \eqref{Eq: Constraint_Interference_term} for successful transmission to preserve the probability of dedicated coverage $\mathrm{P_{cov} \rightarrow P_{cov}^{th}}$ in the network. 
\begin{subequations}\label{Eq: P_min_optimization}
\begin{align}
\mathrm{
\underset{p_{i} \in \mathbf{P}}{\mbox{minimize}}~~ } & \mathrm{\sum_{i \in \mathcal{N}} p_{i} \left( G, \mathbf{X} \right) } \label{Eq: Opt_problem_obj}
\\
\text{s.t.} \quad
& \mathrm{\gamma_{i}=\frac{{g}_i {p}_i  r_{i}^{-\alpha}}{\mathcal{I}+\sigma_{n,i}^2} \geq \gamma_{i}^{th}; \quad \forall i \in \mathcal{N}}
\label{Eq: Constraint_SINR}
\\
& \mathrm{\mathcal{I} = \sum_{j \in \mathcal{N} \setminus i} \sum_{\mathbf{x} \in \mathbf{\Phi}_{\mathcal{I}_j}} p_{j} g_{\mathbf{x}} \left\| \mathbf{x} - \mathbf{o}_{i} \right\|^{-\alpha}}
\label{Eq: Constraint_Interference_term}
\\
& \mathrm{p^{min} \leq p_{i} \leq p^{max}}
\label{Eq: Constraint_power_domain}
\end{align}
\end{subequations}
where, $\mathrm{g} = \left| \mathbf{g} \right|^2 \in G \subset \mathbb{R} $ and follows an exponential distribution $\mathrm{G \sim exp(\lambda_g)}$.

IoT devices target higher throughput thresholds or proportional SINR $\mathrm{\gamma_{i}}$, to employ higher order modulation schemes, and for better decoding capabilities of the desired signal at the IoT gateway. In the IoT uplink, although independent distance-based power allocation strategies would satisfy the network constraints, following orthogonal space-time-frequency-spreading (STFS) would be more restricted and violated for dense IoT tiers. In addition, some beneficial IoT entities that are closer to the tagged gateway would suppress weaker typical signals from far IoT nodes, following non-cooperative resource allocation techniques, which is called the near-far problem \cite{Near_far_problem}. IoT node-centralized resource allocation protocols are combinatorial and NP-hard despite following linear constraints, providing an exhaustive search of possible permutations, or rendering data-driven pre-trained models would restrict the network dynamics and be challenging computations for low-complexity IoT processors. On the other hand, decentralized resource management frameworks could be used against a lack of awareness, such as incomplete CSI among transceivers; however, this introduces extra latency overhead for additional synchronization through the repetitive physical uplink control channel (PUCCH) \cite{PUCCH_IoT}, and pilot-guided CSI estimations \cite{Pilot_CSI} to initiate the optimization framework. Moreover, it is far more challenging for unified uplink control frameworks to perform in distributed and heterogeneous IoT tiers, demanding a third-party edge computing platform to align dense IoT tiers and to share tier-local data to network-global gateways. Consequently, IoT devices highly demand simpler and interactive dealings with neighboring tiers in online power regulating schemes, thereby satisfying the $\mathrm{\gamma_{i} \geq \gamma_{i}^{th}}$ under realistic uncertainty and interferences.

\section{Game Modeling}
We model the decentralized uplink power selection as a non-cooperative interactive game to make decisions while enhancing the coverage probability of each IoT tier for better quality of service (QoS) and weaker collisions in the network. IoT entities, including the surrounding objects, are spatially dynamic in each realization and are scattered independently across the network. Accordingly, it is not practicable to form a complete and perfect resource allocation game due to a lack of knowledge of networking parameters. 

\subsection{Imperfect Game Formation}
In this section, we represent the strategic form of the non-cooperative game-theoretic model: a triplet \eqref{Eq: Game_model_imperfect} to solve the multi-objective transmit power minimization problem \eqref{Eq: P_min_optimization}, addressing user interactions among heterogeneous IoT tiers.
\begin{align}
    \mathrm{
    \mathcal{G} \triangleq \Big\langle \mathcal{N}, \{ \mathcal{S}_{i}, \mathcal{U}_{i} \}_{i \in \mathcal{N}} \Big\rangle
    }
    \label{Eq: Game_model_imperfect}
\end{align}
Generally, an IoT player $\mathrm{i \in \mathcal{N}}$ with conflicting interests competes and selects the best power action from a finite set of strategies $\mathrm{\mathcal{S}_{i}}$ in a basic game model, and the cross strategy space is $\mathrm{\mathcal{S}=\times_{i \in \mathcal{N}} \mathcal{S}_i}$. Then, the utility or payoff function $\mathrm{\mathcal{U}_{i} \rightarrow \gamma_{i}}$ with $\mathrm{\mathcal{U} = \{ Pc_{1}, \dots, Pc_{N}  \}}$, and $\mathrm{\gamma_i:\mathcal{S}\rightarrow\mathbb{R}}$ for each player $\mathrm{i}$ measures the degree of coverage satisfaction of the combination of power choices emitted by IoT players, $\mathrm{\mathcal{S}_i \rightarrow p_i}$. The opponents’ strategy profile is $\mathrm{\mathcal{S}_{\setminus i}=\{ \mathcal{S}_1,\dots,\mathcal{S}_{i-1},\mathcal{S}_{i+1},\dots,\mathcal{S}_N \}}$ and mapped to the utility space $\mathrm{\mathcal{U}_{\setminus i}=\{ \mathcal{U}_1,\dots,\mathcal{U}_{i-1},\mathcal{U}_{i+1},\dots,\mathcal{U}_N \}}$. The game model is strictly tied with the multi-objective performance metric captured by the per-player specific channel throughput $\mathrm{\gamma_{i} (\mathcal{S})}$. Moreover, each IoT player $\mathrm{i}$ has partial freedom $(\mathrm{s_{i} \in \mathcal{S_{i}}})$ over the transmit power variable, tied with a distributed optimization framework, making independent decisions to follow some common given rules in the game. Hence, each agent is restricted to be aware of every stage of the game with the exact immediate action set taken by opponents, violating a perfect gaming environment. 
\begin{figure*}[t]
\centering
$\begin{array}{cc}
\includegraphics[width=0.4\textwidth, trim={0mm 1mm 1mm 3mm},clip]{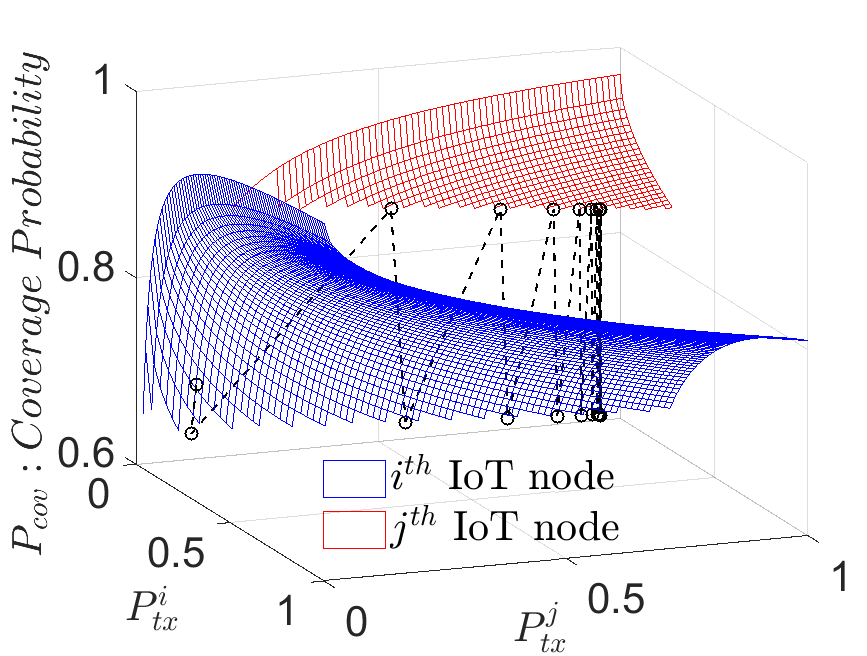} 
&
\includegraphics[width=0.4\textwidth]{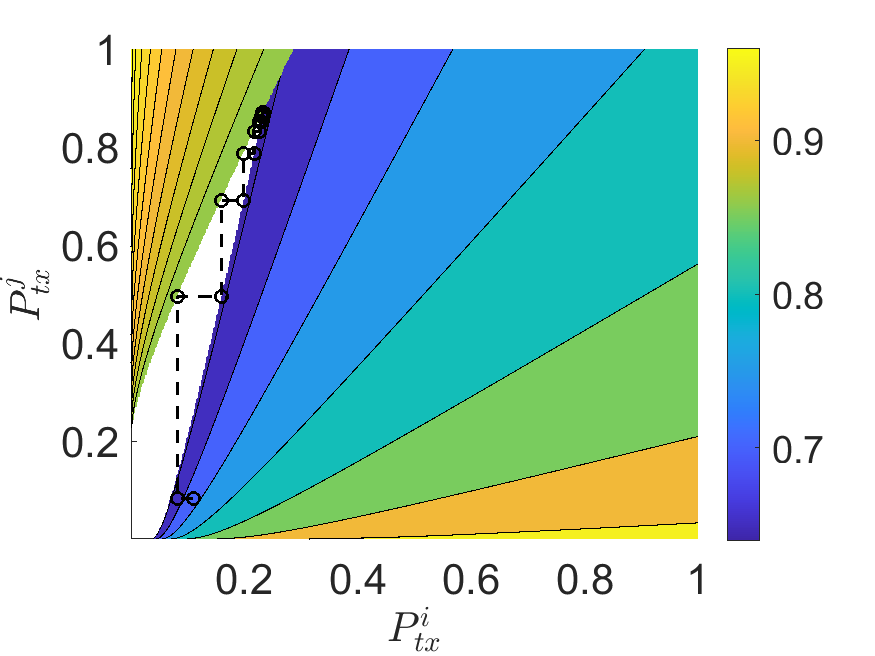}  \\
\mbox{({\textit{a}}) $3D$ Schematic} &   \mbox{({\textit{b}}) $2D$ Schematic}\\
\end{array}$ 
\caption{Visualization of normalized transmit power strategy convergence pattern toward the equilibrium for $2$ IoT nodes, named $i$ and $j$, with desired CSI: $ \mathrm{r_{i} = 63}$ m, $ \mathrm{r_{j} =82 }$ m, $\mathrm{\lambda_{g}=1}$, and intereference CSI: $\mathrm{\lambda=10^{-5}}$ for $\mathbf{x} \in \mathbf{\Phi}_{\mathcal{I}_j}$, $\mathrm{\alpha=4}$, $\mathrm{\sigma_{n}^{2} = -90}$ dB satisfying the given channel throughput threshold of $\mathrm{\gamma^{th}= 0}$ dB on coverage probabilities $\mathrm{P_{cov}^{i}=0.65}$  and $\mathrm{P_{cov}^{i}=0.85}$. 
}
\label{Fig: Imperfect_game}
\end{figure*}
Therefore, all IoT nodes are emitting signals simultaneously in an imperfect game model, energized by the local best power strategy without being informed of the opponent's choice, and cannot change the actions that have already been taken after observing the coverage performance of the network in reality. Consequently, IoT devices are motivated to take spatial advantage while selecting local power optimals under a near-far game model, and it is more beneficial for nodes positioned closer to the IoT gateway and vice versa. Although the local power optimals ride the rival to satisfy the given channel throughput threshold vector for users with higher channel gains, that would penalize the weaker received signals conveying from far IoT users but connected to as the desired. This phenomena is identified as taking striclty dominated strategy over mutually-independent competitors as the best local decisions and creates unnecessary conflicts among IoT transmissions. 

First, we play an imperfect game model in which IoT decision makers are unaware of opponents' strategies but possess perfect CSI in each realization, and we explain optimal navigation to gain a clear intuition for the proposed algorithm. This reflects a complete CSI game \eqref{Eq: Game_model_imperfect} that is aware of exact locations $\mathrm{\mathbf{x} \in \mathbf{X}}$ and non-line-of-sight uncertainties $\mathrm{\mathbf{g} \in \mathbf{G}}$, yet cannot observe or predict the post-transmission power selections $(\mathrm{\mathcal{S}_{j} \rightarrow p_{j}})_{\forall j}$ of opponents. The coverage probability $\mathrm{P_{cov}} \rightarrow \mathrm{\mathcal{U}}$ for a given channel throughput threshold $\mathrm{\gamma^{th}}$ is the utility reward for each desired player. According to \figurename~\ref{Fig: Imperfect_game}(a) network coverage utility is a continuous and concave objective for each desired player power action $\mathrm{p_{i}}$ against the strategy of the opponents $\mathrm{\times_{j \in \mathcal{N}\setminus i}\mathcal{S}_{j}}$. Then, the (Nash) equilibrium $\mathbf{p^{*}}$ of monotone increasing n-node game such that,
\begin{multline*}
\mathrm{P_{cov}^{i}(\mathbf{p}^{*}) \;=\; \max_{x_{i}} \{ \gamma_{i}\big(p_{1}^{*},\ldots,p_{i-1}^{*},x_{i},p_{i+1}^{*},\ldots,p_{n}^{*}\big)} \\
\mathrm{\ \big|\ (p_{1}^{*},\ldots,x_{i},\ldots, p_{n}^{*}) \in\mathbb{R} \}; \quad \forall i \in \mathcal{N}}
\end{multline*}
All desired IoT players have an advantage to draw the complete map of the game, contrary to the opponent tiers, and then could explore the stability region toward the optimal point. There is no motivated player who has the capability for unilateral deviation from the power strategy at the equilibrium $\mathrm{\mathbf{p^{*}}}$ to gain self-coverage probability in the tier, and that would prove the well-known Bayesian game theory lemma, i.e. \emph{the property of Existence \& Uniqueness}, (see proof in \cite{exist_n_uniquness_NE_proof}) provided the per-player payoff $\mathrm{P_{cov}^{i}(p_i,\mathbf{p}_{-i})}$ is continuous and quasi-concave in $\mathrm{p_i}$ over the compact strategy set $\mathrm{[p^{min},p^{max}]}$, \emph{(proof: Appendix \ref{appendix: cov_prob_concav_interaction} existence, by the Debreu--Glicksberg--Fan theorem, \cite{NE_existance_Debreu_Gerard_theorem})}, and the best-response map is a standard interference function in the sense of Yates, i.e. positive, monotone, and scalable (uniqueness). \rnote{Existence/uniqueness was asserted by citation only; the added conditions must be verified for the coverage-probability payoff (not the usual concave SINR utility).}

\figurename~\ref{Fig: Imperfect_game}(b) depicts the graphical illustration of capturing user interactions among each IoT player on the contour objectives \emph{(two-dimensional)} projected from coverage utilities under $0$ dB channel throughput threshold margin. This game is validated for heterogeneous IoT tiers targeting different coverage probabilities $\mathrm{P_{cov}^{th}}$ as shown in the 3D schematic or explained on the contour surface compared to the colorbar. In the operation, each desired IoT node draws the best instantaneous power response given competitors' authorized strategy set on CSI to play the imperfect game independently, eliminating the strategy-dominated actions towards the pinpointed intersection. This iterative power updating pushes the initial and random power set into the global optimum of the optimization problem \eqref{Eq: P_min_optimization} in the concave setting, stabilizing the achievement of successive targets $\mathrm{\{\gamma^{th}\}}$. As a result, repeated best-response reasoning over this fixed utility game guides all non-cooperative players toward the same equilibrium outcome characterized by the shared payoff structure.

\subsection{Incomplete Game Formation}
Nevertheless, the real-world desired uplink resource assignment problem cannot be considered as an imperfect game formation, and applying a direct iterated protocol to find the optimum allocation is hindered by the incompleteness of neighbour positions $\mathrm{\mathbf{x}_{j}}$ and CSIs $\mathrm{\mathbf{g}_{j}}$. Consequently, the new Bayesian game model is given by \eqref{Eq: Bayesian_Game_model},
\begin{align}
    \mathrm{
        \mathcal{G} \triangleq \Big\langle \mathcal{N}, \mathcal{T}, \phi_{\mathcal{T}}, \{ \mathcal{S}_{i}, \mathcal{U}_{i} \}_{i \in \mathcal{N}} \Big\rangle
        }
        \label{Eq: Bayesian_Game_model}
\end{align}
where $\mathcal{T}$ is the set of states (type $\mathrm{t_{j} \in \mathcal{T}}$), set of IoT players which is used to model unknown parameters such as locations and NLOS small-scale fadings of opponent IoT devices, $\mathrm{t_{j} = \{ \mathbf{g}_{j} \cup \mathbf{x}_{j} \} }$. Equivalently, with individual type spaces $\mathrm{\mathcal{T}_i}$, the joint type space satisfies $\mathrm{\mathcal{T}=\times_{i \in \mathcal{N}} \mathcal{T}_i}$ (i.e., $\mathrm{\{\times_{i \in \mathcal{N}} \{ \mathbf{g}_{i} \cup \mathbf{x}_{i} \}\}=\mathcal{T}; \quad \forall i \in \mathcal{N}}$). The incomplete type realizations are scattered on the given common prior-probability distribution (i.e., $\mathrm{\phi_{\mathcal{T}}}$) shared by all IoT entities in the network is derived as the joint probability density function of $\mathrm{\{G \sim exp(\lambda_{g}) \cup \mathbf{x} \in \mathbf{\Phi}_{\mathcal{I}} \}}$ and satisfy $\mathrm{\int_{-\infty}^{\infty} \phi_{\mathcal{T}} \, dt_i~=~1;\ \forall i \in \mathcal{N}}$. In addition to the strategy imperfection $\mathrm{\mathcal{S}}$ of the game, the channel throughput of the desired node $\mathrm{i}$ is no longer deterministic $\mathrm{\gamma_{i}: \mathcal{S} \times \mathcal{T} \rightarrow \mathbb{R}}$, and generates a random coverage utility space $\mathrm{\mathcal{U} \rightarrow P_{cov}}$. Although the desired IoT node knows about self-CSIs to the tagged IoT gateway, it would not be possible to observe neighbor CSIs due to the highly dynamic and heterogeneous nature of each IoT tier. This blocks direct iteration toward the global optimum and challenges the consideration of stochastic utility surfaces for opponent IoT devices from the desired node's perspective.

\section{Epistemology for Bayesian Games} \label{Sec: Epistemology_for_Bayesian_Games}
Typical IoT players $\mathrm{i \in \mathcal{N}}$ are encouraged to make rational decisions instead of relying on best response exploration with respect to the opponent's strategy set $\mathrm{\mathcal{S}_{j}}$. The best rational response based on beliefs $\mathrm{s^{*}\!\in\!\mathcal{S}_{i}}$, with $\mathrm{Pr(s_{i}\!\leftarrow\! t_{i}: \{\mathbf{g_{i}}\cup\mathbf{x_{i}}\})\!\in\!\phi_{\mathcal{T}}: f_{\mathbf{G,X}}(\mathbf{g,x}); \quad \forall s\!\in\!\mathcal{S}_{i}}$, is
\begin{align} \label{Eq: rational_best_response}
    \mathrm{
    \sum_{s_{-i}\in \mathcal{S}_{-i}} Pr(s_{-i}).U_{i}\left( s^{*},s_{-i} \right) \geq \sum_{s_{-i}\in \mathcal{S}_{-i}} Pr(s_{-i}).U_{i}\left( s,s_{-i} \right)
    }
\end{align}
The fundamental rational decision-making approach in \eqref{Eq: rational_best_response} demands an expectation-based mechanism for each desired node $\mathrm{i}$ to deal with probabilistic utility spaces over interfering neighbors. 
\subsection{Utility Expectation: Coverage Probability}
In the $\mathrm{i}$-th desired tier, the typical IoT user $\mathrm{i}$ with random distance $\mathrm{r_{i}}$ is assigned to the IoT gateway at $\mathrm{\textbf{o}_{i}}$ located in the distributed IoT network. Since each IoT node is connected to the closest gateway, and no other tier gateways are closer than $\mathrm{r_{i}}$, it implies that all neighbor tier interferences must be farther than $\mathrm{r_{i}}$. Therefore, the statistical $\mathrm{r_{i}}$ distance distribution is initiated by $\mathrm{\Pr(r \geq r_{i}) = \Pr[\text{No Gateway closer than $\mathrm{r_{i}}$}]}$ and generalizes the Rayleigh distribution, which is differentiated from the null probability of the 2D PPP \cite{Distance_distn_Rayl_PPP}, and the PDF is given by
    $\mathrm{f_{R}(r_{i}) = 2 \pi \lambda r_{i} e^{-2 \pi \lambda r_{i}^2}; \quad r_{i} \geq 0}$.
The maximum boundary $\mathrm{\bar{R}}$ for each IoT user's dynamics is then derived from the conditional PDF, which yields the truncated version of the Rayleigh distribution as follows.
\begin{align}\label{Eq: Rayl_PDF_PPP_truncated}
    \mathrm{f_{R}(r_{i}|\bar{R}) = \frac{2 \pi \lambda r_{i} e^{- \pi \lambda r_{i}^2}}{1 - e^{-\pi \lambda \bar{R}^2}}; \quad 0 \leq r_{i} \leq \bar{R}}
\end{align}
Then, the coverage probability of the IoT node $\mathrm{i}$ is defined as $\mathrm{P_{cov}^{i} \triangleq \Pr(\gamma_{i} \geq \gamma^{th})}$, which is the exact complementary cumulative distribution function (CCDF) of the continuous SINR function $\mathrm{\gamma_{i}}$ over the heterogeneous IoT tiers and from the CDF of $\mathrm{\Pr(\gamma_{i} \leq \gamma^{th})}$. This elaborates the probability of achieving the given channel throughput threshold target $\mathrm{\gamma^{th}}$ by a typical IoT device $\mathrm{i}$ or an average fraction of the network \emph{(desired tier)} that fulfills the on-demand coverage, i.e.
\begin{align} \label{Eq: covergae_probability_initialized}
    \begin{split}
        \mathrm{
        P_{cov}^{i}} &= \mathrm{\mathbb{E}_{R}\left[ \Pr \left( \gamma_{i} \geq \gamma^{th} \big| r=r_{i} \right) \right]}\\
        &= \mathrm{ \int_{0}^{\bar{R}} \Pr \left( \gamma_{i} \geq \gamma^{th} \big| r_{i} \right) . f_{R}(r_{i}|\bar{R}) dr_{i}
        }
    \end{split}
\end{align}
Then, the expected payoff function for each desired IoT node $\mathrm{i}$ is given, by \emph{(proof: Appendix \ref{appendix: coverage_prob}, inserting \eqref{Eq: inner_prob_expectation} into \eqref{Eq: coverage_prob_integral}, then substituting the Laplacian interference term derived as \eqref{Eq: intf_charac} in the appendix part \ref{appendix: Laplace_interfernce_characetrization} with $\mathrm{\zeta = \lambda_g \gamma^{th} {p}_i^{-1}  r_{i}^{\alpha}}$ and simplified.)} 
\begin{align} \label{Eq: covergae_probability_derived}
    \mathrm{P_{cov}^{i} = \frac{2 \pi \lambda}{1 - e^{-\pi \lambda \bar{R}^2}} \bigintssss_{0}^{\bar{R}} r_{i} e^{- \left(\pi \lambda \Psi r_{i}^2 + \lambda_g \gamma^{th} {p}_i^{-1}  r_{i}^{\alpha} \sigma_{n,i}^2 \right)} dr_{i}}
\end{align}
where, $\mathrm{\Psi = 1 + \left(\gamma^{th} {p}_i^{-1} \right)^{2/\alpha} \Upsilon(\alpha) \mathbb{M}_{\frac{2}{\alpha}} \hspace{-1mm} \left[ p_{j} \right]}$ including the improper integral representation of $\mathrm{\Upsilon(\alpha)}$, \cite{imprpoer_integral} and the $\mathrm{(2/\alpha)}$-th order moment generation function (MGF) of the opponent (interference) IoT uplink power vector, i.e., $\mathrm{\mathbb{M}_{\frac{2}{\alpha}} \hspace{-1mm} \left[ p_{j} \right]}$. Before taking the measure about firing strength, it is natural to reason about interference IoT devices' not only their choices but also beliefs of the opponent IoT players. In the realm of epistemology, making beleifs is the key strategy to build the map of the game against to opponent players to take the best decision that would impact for stability of the entire system. However, we should follow a systematic way to build the beliefs hierarchy without loss of generality of stochastic events in the network and avoiding mis-beleifs spaces that would create local equilibriums in the desired game \eqref{Eq: Bayesian_Game_model}. We propose the resource allocation algorithm following the comparison of beliefs with opponent IoT players (inter) in each layer of the hierarchy and the self-awareness of the beliefs that have taken (intra) to hop among layers to build the entire belief space. Hence, players' beliefs and rationality shape strategic choices in the epistemic settings that are quantified in the following part.
\subsection{Inter-epistemic beliefs}
The proposed epistemic design identifies each stochastic state of all interfering IoT players and builds the conditional belief hierarchy by transforming the initial hypothesis into evidence. Here, the ideology of inter-epistemic beliefs represents: the actual desired IoT node $\mathrm{i}$ \emph{(author-desired IoT node)} generates the set of hypothesis over each immediate opponent IoT player $\mathrm{k}$, treating it as the desired \emph{(epistemic-desired IoT node)}, while the author-desired node $\mathrm{i}$ is in the interference pool, and yields the hypothesis set in \eqref{Eq: inter_epist_H}. The author-desired IoT player $\mathrm{i}$ is aware of self-CSI and initializes the power action $\mathrm{p_{i}^{l}}$ in the self-comparison stage $\mathrm{l}$. Then node $\mathrm{i}$ anticipates possible transmit power actions $\mathrm{p_{k}^{(m)}}$ for the emulated node $\mathrm{k}$ thereby, exceeding the given channel throughput threshold $\mathrm{\gamma_{k}^{th}}$. The rest of the opponent IoT players' power actions are taken from the prior epistemic stage $\mathrm{(m\!-\!1)}$, and the posterior $\mathrm{(m)}$ probabilistic transition to build the average of the degree of belief, and generalize the hypothesis set, which is given by 
\begin{align}
    \mathrm{\mathcal{H}_{k, inter}^{(l,m)}} \hspace{-2mm} &= \hspace{-1mm} \mathrm{\left\{\hspace{-0.5mm} \mathbb{E} \hspace{-1mm} \left[ \Pr \hspace{-1mm} \left( \gamma_{k} \hspace{-1mm} \left( p_{k}^{(m)} \right) \hspace{-1mm} \geq \hspace{-1mm} \gamma_{k}^{th} \Big| \hspace{-1mm} \left\langle \mathbf{x}_{i}, p_{i}^{(l)}, p_{j \in \mathcal{N}\setminus \{i,k\}}^{(m-1)} \right\rangle \hspace{-0.5mm} \right) \hspace{-0.5mm} \right] \hspace{-0.5mm} \right\}_{p_{k}^{(m)} \in \textbf{P}}} 
    \label{Eq: inter_epist_H}
\end{align}
Therefore, we can reformulate the hypothetical coverage probability vector as shown in \eqref{Eq: cov_prob_inter_epistemic} for the inter-epistemic belief strategy employing the derived fundamental formula in \eqref{Eq: covergae_probability_derived}. 
\begin{align} \label{Eq: cov_prob_inter_epistemic}
    \mathrm{P_{cov}^{k} \big|^{(m)} \hspace{-1mm} = \hspace{-1mm} \frac{2 \pi \lambda}{1 - e^{-\pi \lambda \bar{R}^2}} \hspace{-1mm} \bigintssss_{0}^{\bar{R}} \hspace{-2mm} r_{k} e^{- \left(\pi \lambda \Psi_{k}^{(l,m)} r_{k}^2 + \frac{\lambda_g \gamma^{th} r_{k}^{\alpha} \sigma_{n,k}^2}{{p}_k^{(m)}} \right)} dr_{k}}
\end{align}
s.t, $\mathrm{\Psi^{(l,m)}_{k} = 1 + \left( \frac{\gamma^{th}}{p_{k}^{(m)} } \right)^{2/\alpha} \hspace{-2mm} \Upsilon(\alpha) \mathbb{M}_{\frac{2}{\alpha}} \hspace{-1mm} \left[ p_{i}^{l}, \left\langle p_{j \in \mathcal{N}\setminus \{i,k\}}^{(m-1)} \right\rangle\right]}$.
\begin{figure}[t]
\centering
\includegraphics[width=0.9\columnwidth, trim={0mm 0mm 0mm 0mm},clip]{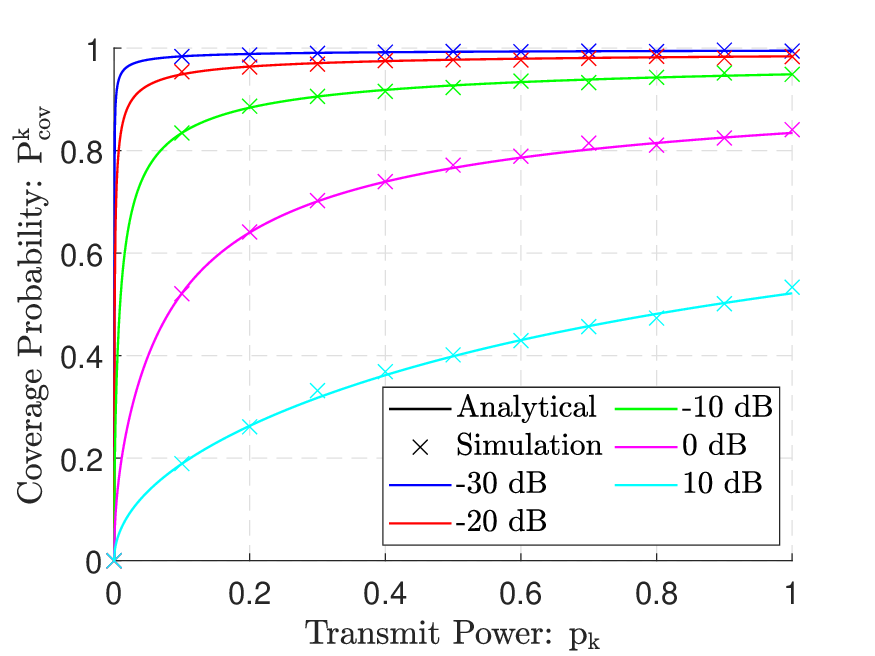}
\caption{Comparison of analytical and simulation (number of iterations: $\mathrm{I_{inter}=5000}$) coverage probability $\mathrm{\left(P_{cov}^{k} \text{ Vs } p_{k}\right)}$ results, of the replicated IoT tier $\mathrm{k \in \mathcal{N}\setminus i}$, against the transmit power strength $\mathrm{p_{k}}$ of epistemic-desired node in the inter-epistemic stage for the set of channel throughput threshold $\mathrm{\gamma^{th}} = \{-30, -20, -10, 0, 10\}$ dB, under parameter settings: $\mathrm{\lambda_{g}=1}$, $\mathrm{\lambda=-50}$ dB, $\mathrm{\alpha=4}$, $\mathrm{\sigma_{n}^{2} = -90}$ dB, $\mathrm{\bar{R}=100}$ m.}
\label{Fig: P_coverage_vs_P_k_inter_epistemic}
\end{figure}
\figurename~\ref{Fig: P_coverage_vs_P_k_inter_epistemic} represents the hypothetical coverage beliefs that satisfy the given channel throughput bound and formulated in \eqref{Eq: cov_prob_inter_epistemic} of the epistemic-desired IoT device $\mathrm{k}$ with respect to the firing strength. This represents a single inter-epistemic transition realization from $\mathrm{(m\!-\!1) \rightarrow (m)}$ at the $\mathrm{l}$-th state. It is evident that the coverage probability $\mathrm{P_{cov}^{k}}$ of the $\mathrm{k}$-th IoT device increases monotonically with uplink transmit power vector $\mathrm{p_{k}}$, while lower throughput threshold margins lead to an upward shift for the network coverage. Moreover, the cross points indicate the realistic expected coverage performances (simulation) for given transmit power vectors and lie on the generalized analytical inter-epistemic beliefs strategies \eqref{Eq: inter_epist_H}. This motivates the $\mathrm{i}$-th author-desired IoT node to analyze the belief space for further evaluation aligned with practical network dynamics.

Thereafter, the author-desired player $\mathrm{i}$ recognizes the best instantaneous response $\mathrm{\left(p_{k}^{(m)}\right)^{*}}$ \emph{there exist} evidence $\mathrm{\mathcal{E}_{(k, inter)}^{(l,m)}}$ at state $\mathrm{(l,m)}$, to maximize the degree of belief over the hypotheses, preserving the coverage probability $\mathrm{P_{cov}^{th}}$ of the $\mathrm{k}$-th epistemic-desired IoT tier. This elaborates an \emph{inter-epistemic} chain such that $\mathrm{i}$-th player updates for all competitive opponents; hence, the inter-epistemic transition from $\mathrm{(m\!-\!1)}$ to $\mathrm{(m)}$ for a fixed intra-stage $l$, is
\begin{align}
   \mathrm{\mathcal{E}_{(k, inter)}^{(l,m)} = \exists \hspace{1mm} \mathcal{H}_{(k, inter)}^{(l,m)}\hspace{-1mm}\left[\left( p_{k}^{(m)} \right)^{*}, p_{i}^{(l)}, p_{j \in \mathcal{N}\setminus \{i,k\}}^{(m-1)} \right] \geq P_{cov}^{th}} \label{Eq: inter_epist_E}
\end{align}
Although the author-desired player $\mathrm{i}$ builds the belief layer against the opponent IoT players following inter-epistemic transitions in state $\mathrm{l}$, it is not sufficient to utilize the instantaneous transmit power level $\mathrm{p_{i}^{(l,\mathcal{M}_{l})}}$ in the uplink. That should be further verified whether the optimal stage that would stabilize the game or have further potential to deviate for upper levels if $\left(\gamma_{i}^{(l,\mathcal{M}_{l})} \geq \gamma^{th}\right)$ or lower targets if $\left(\gamma_{j}^{(l,\mathcal{M}_{l})} < \gamma^{th} \right)_{j \in \mathcal{N} \setminus i}$, which is explained in the following section.

\subsection{Intra-epistemic beliefs}
This represents the inter-epistemic layer transitions $\mathrm{(l \rightarrow \mathcal{L}, \mathcal{M}_{l})}$ by the $\mathrm{i}$-th author-node, after processing against all opponent IoT players as epistemic-desired with respect to the existing transmit power level $\mathrm{p_{i}^{(l)}}$. Therefore, the author-desired IoT player makes an intrinsic evaluation about the current power action that has taken $\mathrm{p_{i}^{(l)}}$ and updates it toward the global equilibrium $\mathrm{p_{i}^{(l+1)}}$ in the Bayesian game model. As a result, this sub-optimal mechanism is called an intra-epistemic transition and opens a new belief layer in the hierarchy. Then, the author-desired IoT node $\mathrm{i}$ makes the set of hypothesis in the state $\mathrm{(l+1)}$ exceed the channel throughput limit of $\mathrm{\gamma^{th}}$, over the instantaneous location and the evidence power strategy $\mathrm{p_{j \in \mathcal{N}\setminus \{i\}}^{(l,\mathcal{M}_{l})}}$ of opponents that is filtered from the previous state $\mathrm{(l)}$ is given by,
\begin{align}
    \mathrm{\mathcal{H}_{i, intra}^{(l+1,\mathcal{M}_{l})}} \hspace{-1mm} &= \hspace{-0.5mm} \mathrm{\left\{\hspace{-0.5mm} \mathbb{E} \hspace{-1mm} \left[ \Pr \hspace{-1mm} \left( \gamma_{i} \hspace{-1mm} \left( p_{i}^{(l+1)} \right) \hspace{-1mm} \geq \hspace{-0.5mm} \gamma_{i}^{th} \Big| \hspace{-0.5mm} \left\langle r_{i}, p_{j \in \mathcal{N}\setminus \{i\}}^{(l,\mathcal{M}_{l})} \right\rangle \hspace{-0.5mm} \right) \hspace{-0.5mm} \right] \hspace{-0.5mm} \right\}_{p_{i}^{(l+1)} \in \textbf{P}}} \label{Eq: intra_epist_H}
\end{align}
Here, the expectation of $\mathbb{E}_{\mathcal{I}}$ is limited to consider with respect to the randomness of the opponent interference node only, and the tagged distance $\mathrm{r_{i}}$ from the connected gateway is known to the author-desired IoT node $\mathrm{i}$. Therefore, we can remove the standard integral form for deterministic distance $\mathrm{r_{i}}$ and is simplified under \eqref{Eq: inner_prob_expectation} in the appendix \eqref{appendix: coverage_prob}. Thereafter, the Laplacian interference term is replaced by the derived formula \eqref{Eq: intf_charac} in the appendix \eqref{appendix: Laplace_interfernce_characetrization} to derive the hypothetical coverage probability space \eqref{Eq: cov_prob_intra_epistemic} to the intra-epistemic stage.
\begin{align} \label{Eq: cov_prob_intra_epistemic}
    \mathrm{P_{cov}^{i} \big|^{(l+1)}  =  exp\left(-\pi \lambda \Psi_{i}^{(l+1,\mathcal{M}_{l})} r_{i}^2 - \frac{\lambda_g \gamma^{th} r_{i}^{\alpha} \sigma_{n,i}^2}{{p}_i^{(l+1)}} \right)}
\end{align}
s.t, $\mathrm{\Psi^{(l+1,\mathcal{M}_{l})}_{i} = \left( \frac{\gamma^{th}}{p_{i}^{(l+1)} } \right)^{2/\alpha} \hspace{-1mm} \Upsilon(\alpha) \mathbb{M}_{\frac{2}{\alpha}} \hspace{-1mm} \left[ \left\langle p_{j \in \mathcal{N}\setminus \{i\}}^{(l, \mathcal{M}_{l})} \right\rangle\right]}$.
\begin{figure}[t]
\centering
\includegraphics[width=0.9\columnwidth, trim={0mm 0mm 0mm 0mm},clip]{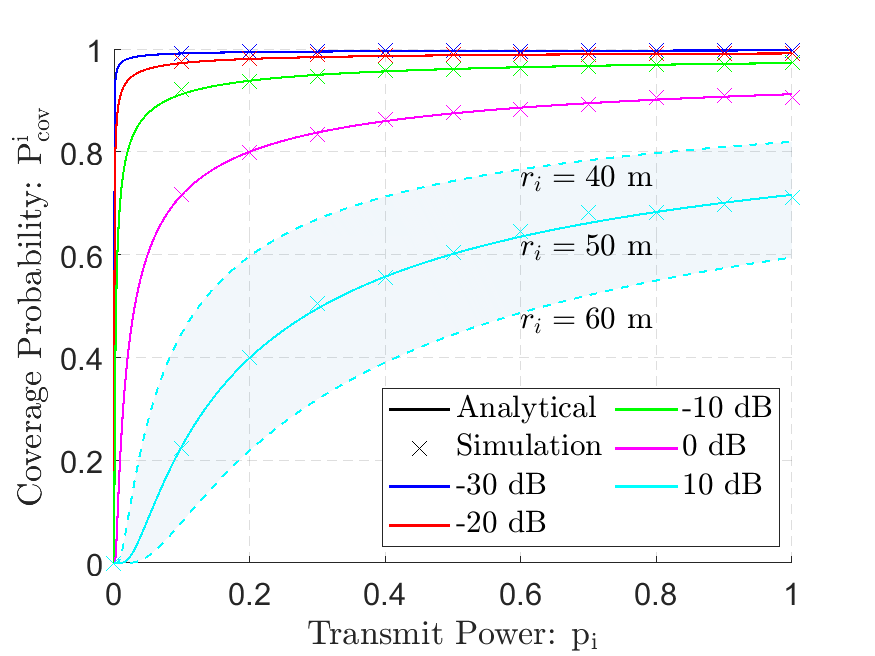}
\caption{Comparison of analytical and simulation (number of iterations: $\mathrm{I_{intra}=5000}$) coverage probability $\mathrm{\left(P_{cov}^{i} \text{ Vs } p_{i}\right)}$ results, of the author-desired IoT tier $\mathrm{i \in \mathcal{N}}$, against the transmit power strength $\mathrm{p_{i}}$ with tagged distance $\mathrm{r_{i}=50}$ m in the intra-epistemic stage for the set of channel throughput threshold $\mathrm{\gamma^{th}} = \{-30, -20, -10, 0, 10\}$ dB, under parameter settings: $\mathrm{\lambda_{g}=1}$, $\mathrm{\lambda=-50}$ dB, $\mathrm{\alpha=4}$, $\mathrm{\sigma_{n}^{2} = -90}$ dB, $\mathrm{\bar{R}=100}$ m.}
\label{Fig: P_coverage_vs_P_i_intra_epistemic}
\end{figure}
Similarly, \figurename~\ref{Fig: P_coverage_vs_P_i_intra_epistemic} shows the analytical and simulation coverage performance $\mathrm{P_{cov}^{i}}$ of the author-IoT device against the firing vector $\mathrm{p_{i}}$. We have validated that the derived hypothetical formula explicitly follows the realistic network coverage performances for distinct throughput boundaries in the intra-epistemic stage as well. More specifically, $\mathrm{P_{cov}^{i}}$ at $\gamma^{th}=10$ dB illustrates that the ability of achieving higher coverage potential by the actual IoT player which are located more closer to the IoT gateway for all around the uplink power strategy, $\mathrm{P_{cov}^{i}|_{r_{i}=40 \text{ m}} > P_{cov}^{i}|_{r_{i}=50 \text{ m}} > P_{cov}^{i}|_{r_{i}=60 \text{ m}}}$. This is a crucial feature for the author-desired IoT node that is the decision maker and aware only of the knowledge of tagged distance $\mathrm{r_{i}}$ to the gateway in the intr-epistemic transition while extracting evidence set \eqref{Eq: intra_epist_E} in the state $\mathrm{(l+1)}$, more intuitively.
\begin{align}
   \mathrm{\mathcal{E}_{(i, intra)}^{(l+1,\mathcal{M}_{l})} = \exists \hspace{1mm} \mathcal{H}_{(i, intra)}^{(l+1,\mathcal{M}_{l})}\hspace{-1mm}\left[\left( p_{i}^{(l+1)} \right)^{*}, p_{j \in \mathcal{N}\setminus \{i\}}^{(l,\mathcal{M}_{l})} \right] \geq P_{cov}^{th}} \label{Eq: intra_epist_E}
\end{align}
Subsequent to the construction of the belief structure, the $\mathrm{i}$-th author-desired IoT node obtains the best power decision $\mathrm{p_{i}^{(l+1)}}$, that \emph{there exists} a hypothesis $\mathrm{\mathcal{H}_{(i, intra)}^{(l+1,\mathcal{M}_{l})}}$ at the existing intra-epistemic stage $\mathrm{(l+1)}$ preserving the prescribed network coverage probability upper bound of $\mathrm{P_{cov}^{th}}$. Consequently, the author-desired player enables a more confident selection of the hypothesis with the highest degree of belief as evidence, i.e., nominated network coverage  $\mathrm{\mathcal{E}_{(i, intra)}^{(l+1,\mathcal{M}_{l})}}$ consuming minimal transmission power for the uplink. Then, we will discuss the intuition behind the confidence of each player to converge toward the global optimum, and the credibility of following Bayesianism to build the inter-intra-epistemic belief chain on top of a common prior.

\subsection{Inter-Intra-epistemic-Rationality}
Each rational player $\mathrm{i}$ builds the epistemic-belif hierarchy $\mathrm{\mathcal{B}_{(l^{*} \times m^{*})}}$ proceeding multi-level iterations up to Nash equilibrium \eqref{Eq: rational_best_response}, located at the inter-intra-epistemic stage $\mathrm{(l^{*},m^{*}) \in (\mathcal{L}, \mathcal{M})}$, where no player can unilaterally improve self-payoff by deviating from the current strategy $\mathrm{\left\{ \left(p_{i} \right)^{*}, p_{j \in \mathcal{N}\setminus \{i\}}^{(l,\mathcal{M}_{l})} \right \}^{(l^{*},m^{*})}}$. We formulate the rational belief on the $\mathrm{i}$-th player about a random event $\mathrm{E}$ conditioned of $\mathrm{F}$ as $\mathrm{B_{i}^{F} = \left \{ w: Max_{\geq i} \left[\Omega_{i}(w) \cap F \right] \subseteq E \right \}}$ where $\mathrm{w = r_{i}}$ contains the partition of state space obtained from $\mathrm{\Omega_{i}(w)}$, \cite{Inter_Intra_epistemic_rational}. The known private information of the subspace $\mathrm{\Omega_{i}(w)}$ steers each player's belief hierarchy toward the Nash power equilibrium at $\mathrm{(l^{*},m^{*})}$, compensating the belief expectation about opponents. Then, the author-desired IoT player $\mathrm{i}$ tailors inter-epistemic belief layer on each epistemic-opponent $\mathrm{k}$, that binds the event $\mathrm{E_{inter} = \mathcal{H}_{k, inter}^{(l,m)}}$ (current hypothsis) conditioned on the established evidence $\mathrm{F_{inter} = \mathcal{E}_{k, inter}^{(l,m-1)}}$ in accordance with \eqref{Eq: inter_epist_H} and \eqref{Eq: inter_epist_E} respectively. Thereafter, intra-epistemic layer-layer belief structure updates hypothesis $\mathrm{E_{intra} = \mathcal{H}_{i, intra}^{(l,m)}}$ following the immediate established evidence $\mathrm{F_{intra} = \mathcal{E}_{i, intra}^{(l+1,m)}}$ as defined in \eqref{Eq: intra_epist_H} and \eqref{Eq: intra_epist_E} respectively.

\begin{figure}[t]
\centering
\includegraphics[width=0.9\columnwidth, trim={0mm 0mm 0mm 0mm},clip]{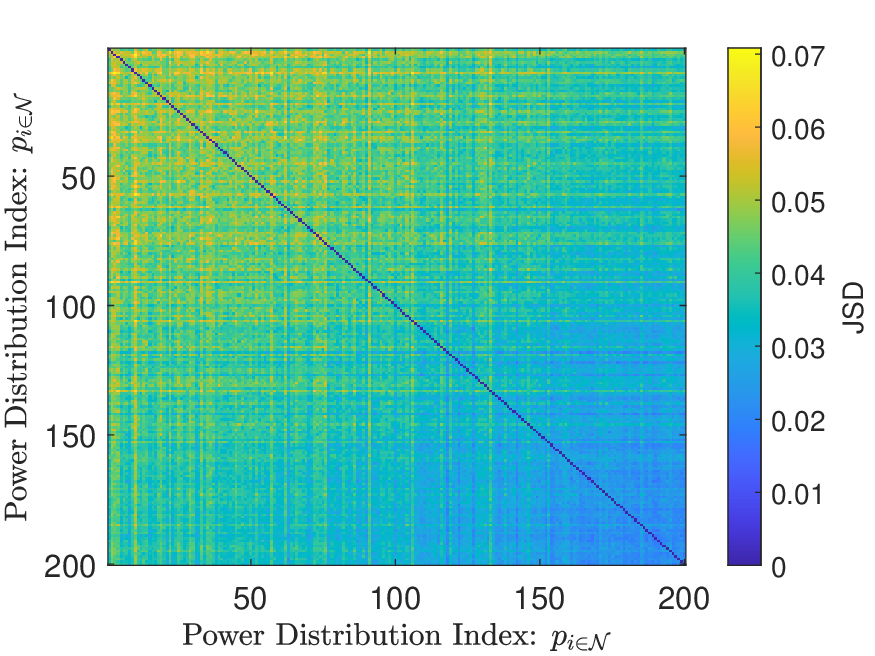}
\caption{Heatmap of the Jensen-Shannon Divergence (JSD) of power profile distributions for each IoT player $\mathrm{i \in \mathcal{N}}$ for $\mathrm{N=200}$, with respect to intra-beleif (diagonal) and inter-beleif (off-diagonal) transmit power vectors that satisfy the channel throughput threshold $\mathrm{\gamma^{th}} = -10$ dB and preserving network coverage $\mathrm{P_{cov} = 80\%}$, under parameter settings: $\mathrm{\lambda_{g}=1}$, $\mathrm{\lambda=-60}$ dB, $\mathrm{\alpha=4}$, $\mathrm{\sigma_{n}^{2} = -90}$ dB, $\mathrm{\bar{R}=120}$ m.}
\label{Fig: JSD_power_action_distn}
\end{figure}
The power matrix $\mathrm{\textbf{P}_{N \times N}}$, where the size of the $\mathrm{N}$-number of IoT players, is updated in each layer $\mathrm{(l,m) \in (\mathcal{L}, \mathcal{M})}$ of the hierarchy of inter-intra-epistemic beliefs. The vector $\mathrm{\langle p_{i} \rangle}_{N \times 1}$ represents the instantaneous power profile of the $\mathrm{i}$-th author-desired IoT player. The diagonal vector $\mathrm{p_{i \times i}}$ of the power matrix $\mathrm{\textbf{P}_{N \times N}}$ consists of actual firing power vectors that have been computed by each author-desired player. Additionally, the off-diagonals represent the power belief profile $\mathrm{p_{i \times j}}$, where $\mathrm{j \in \mathcal{N}\setminus i}$ are for opponents. In the inter-epistemic chain, the author IoT player computes the opponent's potential power vector independently, though each power element would not be aligned to the corresponding opponent identification. However, the characteristics of the entire belief distribution of each author-desired IoT player against opponents would be similar and would maintain a greater correlation. \figurename~\ref{Fig: JSD_power_action_distn} illustrates the heatmap for Jensen-Shannon Divergence (JSD) \cite{JSD} to measure the correlation among each power profile generalized through the inter-intra-epistemic belief hierarchy. Here, the belief distribution deviations are bounded $\mathrm{(0 \leq JSD \leq 1)}$ where the higher correlations are biased toward a lower margin and vice versa. Therefore, all IoT players have constructed the desired and opponent-based power profiles with minimal deviations $\mathrm{(JSD^{max}=0.07)}$ among belief distributions. In addition, the exact diagonal has zero deviation compared to itself, and the overall JSD metric on the colormap sweeps diagonally toward more blueness, while enhancing the belief correlations. Moreover, the color shading is smooth, which implies the consecutive belief updates in the inter-intra-epistemic strategies, and all players are relying on the common prior. Thus, IoT players can address the challenge of incomplete knowledge about opponents to make optimal though, independent decisions on a virtually collaborative platform, referred to as the belief hierarchy.
\section{Proposed Algorithm Overview}
\begin{algorithm}[t]
    \caption{Epistemology-Inspired Bayesian Game Model}
    \label{Alg: Epistemic_BGT_proposed}
    \begin{algorithmic}[1]
        \State Initialize $\mathrm{
        \mathcal{G} \triangleq \Big\langle \mathcal{N}, \mathcal{T}, \phi_{\mathcal{T}}, \{ \mathcal{S}_{i}, \mathcal{U}_{i} \}_{i \in \mathcal{N}} \Big\rangle
        }$ Bayesian game model, \eqref{Eq: Bayesian_Game_model} with network parameters $\mathrm{\Phi_{PPP}(\lambda), \lambda_{g}, \alpha, r_{i \in \mathcal{N}}, \bar{R}, \sigma_{n}^{2}, R_{max}, \gamma^{th}, P_{cov}^{th}}$.
		\For {Author-desired IoT Player $\mathrm{i \in \mathcal{N}}$}
            \State Initialize power profile $\mathrm{\textbf{p}_{vec}^{intial}}$ by author-desired IoT player $\mathrm{i \in \mathcal{N}}$.
            \While{$\mathrm{RMSE\left( p\left[B_{i}^{\mathcal{H}_{i}} \right]  \right) \geq \epsilon \quad \cup \quad iter \leq I^{max}}$} \label{while_loop_NE}
            \For{Random epistemic-desired Player $ \mathrm{k \in \mathcal{N}}$}
                \State Calculate the moment of $\mathrm{\mathbb{M}_{\frac{2}{\alpha}}\left[ \textbf{p}_{vec, \forall j \in \mathcal{N}\setminus i} \right]}$.
                \If {$\mathrm{k \neq i}$} \Comment{Inter-epistemic update}
                    \State \hspace{-2mm}Builds inter-beleif hypothesis layer, \eqref{Eq: inter_epist_H}.
                    \State \hspace{-2mm}Compute optimal $\mathrm{p_{k \in \mathcal{N}\setminus i}}$ following \hspace{-1mm} \eqref{Eq: cov_prob_inter_epistemic}, \hspace{-2mm} \eqref{Eq: inter_epist_E}
                \Else {$\mathrm{\quad ( k = i)}$} \Comment{Intra-epistemic update}
                    \State \hspace{-2mm}Builds intra-beleif hypothesis layer, \eqref{Eq: intra_epist_H}.
                    \State \hspace{-2mm}Compute optimal $\mathrm{p_{i \in \mathcal{N}\setminus k}}$ following \hspace{-1mm} \eqref{Eq: cov_prob_intra_epistemic}, \hspace{-2mm} \eqref{Eq: intra_epist_E}
                \EndIf
                \State Store $\mathrm{p_{i}}, \mathrm{p_{k}}$ in $\mathrm{\textbf{p}_{vec}}$.
            \EndFor
            \State Store $\mathrm{\textbf{p}_{vec}}$ in power strategy matrix, $\mathrm{\textbf{P}_{strategy}}$.
            \State Update the $\mathrm{RMSE\left( p\left[B_{i}^{\mathcal{H}_{i}, now} \right], p\left[B_{i}^{\mathcal{H}_{i}, prev} \right]  \right)}$ \label{RMSE}
            \EndWhile \Comment{Check terminal conditions}
		\EndFor
        \State The diagonal of $\mathrm{\textbf{P}_{strategy}}$ represents the optimal power allocation $\mathrm{\textbf{p}^{*}; \quad \forall i \in \mathcal{N}}$.
    \end{algorithmic}
\end{algorithm}
In this section, we discuss the way to deal with the transmit power minimization problem \eqref{Eq: P_min_optimization} utilizing the proposed epistemology-inspired Bayesian game algorithm. There are a number of IoT devices lying on heterogeneous IoT tiers as shown in \figurename~\ref{Fig: EBGT_System_model} and follows a Poisson point Process, $\mathrm{\Phi_{PPP}}$ with density $\mathrm{\lambda}$. Owing to the higher spectrum resolution, hardware and channel impairments, and device non-linearities, the received signals at the gateway would collide with each other under minimal band gaps, causing frequency offsets and time jitters. Therefore, each IoT device has a self-discipline to regulate firing strength that would preserve the indicated network performance boundaries $\mathrm{\gamma^{th}, P_{cov}^{th}}$ but not exceed, penalizing neighbouring devices that receive signals at the IoT gateway. Therefore, we initialize the Bayesian game model \eqref{Eq: Bayesian_Game_model} that includes all IoT devices in the network, considering interference as opponents. All desired IoT nodes $\mathrm{i \in \mathcal{N}}$ are tagged to the nearest gateway with distance $\mathrm{r_{i}}$ following an exponential distribution and truncated by $\mathrm{\bar{R}}$, \eqref{Eq: Rayl_PDF_PPP_truncated}. 

\begin{figure}[t]
\centering
\includegraphics[width=0.8\columnwidth, trim={2mm 2mm 9mm 6mm},clip]{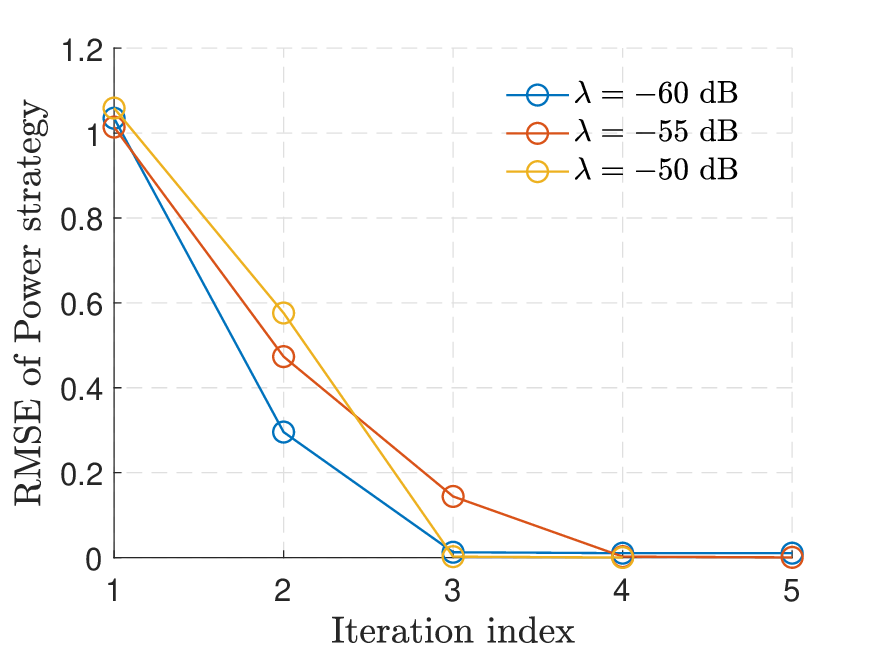}
\caption{Variation of RMSE between two consecutive power belief strategies of IoT players $\mathrm{i \in \mathcal{N}}$, against iteration index illustrating the convergence of the EBGT framework to the global equilibrium, set of PPP mean user densities $\mathrm{\lambda=\{-60, -55, -50\}}$ dB,  under parameter settings: $\mathrm{\gamma^{th}} = 0$ dB, $\mathrm{P_{cov}^{th}} = 0.95$, $\mathrm{\lambda_{g}=1}$, $\mathrm{\alpha=4}$, $\mathrm{\sigma_{n}^{2} = -90}$ dB, $\mathrm{\bar{R}=120}$~m, $\mathrm{Area=10}$~km$^2$.}
\label{Fig: Rmse_iteration}
\end{figure}
The proposed algorithm~\ref{Alg: Epistemic_BGT_proposed} is executed in each IoT node in parallel. We consider the decision maker known as the author-desired IoT player $\mathrm{i \in \mathcal{N}}$, who holds the beliefs about the neighbour opponents under the available common prior distribution. First, the author-desired player $\mathrm{i}$, initializes the transmit power profile $\mathrm{\textbf{p}_{vec}^{intial}}$. Next, the author-desired IoT player starts to explore the Nash equilibrium \eqref{Eq: rational_best_response} of the Bayesian game model under \emph{while loop} at line~\ref{while_loop_NE}, which is terminated with no motivation for unilateral deviations $\mathrm{(\epsilon=-50 dB)}$, $\mathrm{RMSE}$ between the existing and the immediate actions that have been taken. The proposed EBGT algorithm converges to the global equilibrium following the root mean square error (RMSE) between the existing and the immediate power actions, \emph{(line \ref{RMSE})} as shown in \figurename~\ref{Fig: Rmse_iteration}.

Then, the author-desired IoT player obtains the $\mathrm{\left(\frac{2}{\alpha}\right)th}$ order moment of the power vector $\mathrm{\textbf{p}_{vec, \forall j \in \mathcal{N}\setminus i}}$ except the power element $\mathrm{p_{i}}$ itself. Now, the author-desired IoT node $\mathrm{i( \neq k)}$ considers each sequential interference entity $\mathrm{k \in \mathcal{N}\setminus i}$ as the acting desired (epistemic) tailoring inter-epistemic belief layers \eqref{Eq: inter_epist_H} and secure sub-optimal strategy, $\mathrm{p_{k \in \mathcal{N}\setminus i}}$. This is a $\mathrm{k}$-th node directional linear programming problem that delivers a strictly dominated strategy by each node $\mathrm{k}$ and steers on the author's action $\mathrm{p_{i}}$, \eqref{Eq: cov_prob_inter_epistemic}, \eqref{Eq: inter_epist_E}. On the other hand, the intra-epistemic belief layer \eqref{Eq: intra_epist_H} is re-updated by the author-IoT player $\mathrm{i(=k)}$ and computes the sub-optimal equilibrium stage $\mathrm{p_{i \in \mathcal{N} \setminus k}}$ on top of the opponents' best strategy vector $\mathrm{\textbf{p}_{k \in \mathcal{N}\setminus i}}$, \eqref{Eq: cov_prob_intra_epistemic}, \eqref{Eq: intra_epist_E}. Ultimately, both the actual and the belief power strategies are adapted and stored in a vector called $\mathrm{\textbf{p}_{vec}}$, then stacked in the power matrix $\mathrm{\textbf{P}_{strategy}}$ following each IoT device in the network. This metric represents the inter-intra-epistemic belief hierarchy as graphically illustrated in \figurename~\ref{Fig: JSD_power_action_distn}, and the diagonal represents the equilibrium state of realized firing-power $\mathrm{\textbf{p}^{*}; \quad \forall i \in \mathcal{N}}$.    
\begin{algorithm}[t]
    \caption{Monte-Carlo Validation of Proposed Mechanism}
    \label{Alg: Monte_Carlo_Validation}
    \begin{algorithmic}[1]
        \State Initialize the IoT network parameters $\mathrm{\Phi_{PPP}(\lambda), \lambda_{g}, \alpha, r_{i \in \mathcal{N}}, \sigma_{n}^{2}, R_{max}, \gamma^{th}, P_{cov}^{th}}$.
        \For {The author-desired IoT Player $\mathrm{i \in \mathcal{N}}$}
            \State Extract optimal transmit power $\mathrm{p_{i}^{*} \in \textbf{p}^{*}}$ for node-$\mathrm{i}$.
            \State Make $\mathrm{Counter = 0}$.
            \While{$\mathrm{iter \leq I_{MC}^{max}}$}
                \State Pick the Channel fading gain: $\mathrm{|\textbf{g}|^{2} \in G \sim exp(\lambda_{g})}$.
                \State Random interference distribution: $\mathrm{r_{j}:\Phi_{PPP}(\lambda)}$.
                \State Interference power actions: $\mathrm{p_{j}^{*} \in \textbf{p}^{*}}; \quad \forall j \in \mathcal{N} \setminus i$.
                \State Compute SINR: $\mathrm{\gamma_{i}}$, \eqref{Eq: Constraint_SINR}, \eqref{Eq: Constraint_Interference_term}.
                \If{$\mathrm{\gamma_{i} \geq \gamma^{th}}$}
                    \State $\mathrm{Counter = Counter + 1}$.
                \EndIf
            \EndWhile
            \State $\mathrm{P_{cov}^{i} = \left(Counter/I_{MC}^{max} \right) \rightarrow P_{cov}^{th}}$.
        \EndFor
    \end{algorithmic}
\end{algorithm}

We validate the optimum transmit power vector $\mathrm{\textbf{p}^{*}}$ obtained from the proposed epistemology-based Bayesian game theory algorithm~\ref{Alg: Epistemic_BGT_proposed} that functions on the Monte-Carlo simulation algorithm~\ref{Alg: Monte_Carlo_Validation} under realistic constraints. First, we initialize the metrics of the networks and consider that the author-desired IoT device $\mathrm{i \in \mathcal{N}}$ is connected to the nearest gateway located at the origin $\mathrm{\textbf{o}_{i} \in \textbf{O} \subset \mathbb{C}^{2}}$. Then, the desired node extracts the typical transmit power $\mathrm{p_{i}^{*} \in \textbf{p}^{*} }$, and the algorithm~\ref{Alg: Monte_Carlo_Validation} initializes the $\mathrm{Counter = 0}$ to measure the network coverage capabilities. In each Monte-Carlo iteration, there are random interference realizations, $\mathrm{r_{j}:\Phi_{PPP}(\lambda)}$ sustained with different small-scale path fading gains, $\mathrm{|\textbf{g}|^{2} \in G \sim exp(\lambda_{g})}$. Each IoT device $\mathrm{i \in \mathcal{N}}$  utilize the optimal power strategies $\mathrm{\textbf{p}^{*}}$ to communicate with the connected gateway and indicates the receiver SINR $\mathrm{\gamma_{i}}$. The algorithm counts the number of successive terms that satisfy the nominated $\mathrm{\gamma^{th}}$ bound and the finalized average reveals the coverage probability $\mathrm{P_{cov}^{i}; \quad \forall i \in \mathcal{N}}$ that obeys to the $\mathrm{P_{cov}^{th}}$. 

\section{Results and Discussion}
In this section, we present the simulation results of the proposed epistemology-inspired Bayesian game theory (EBGT) and validate the network performance through the Monte-Carlo evaluation metric that aligns with realistic wireless network parameters. There are $\mathrm{N=150}$ IoT tiers distributed in the network area of $\mathrm{10}$ km$^2$. Each desired IoT gateway radiates the maximum coverage distance $\mathrm{\bar{R} = 120}$ m to the tagged IoT nodes following a homogeneous PPP with density $\lambda$ in dBs in a circular region. The uplink is attenuated with distance to the standard power-path loss propagation model under the path loss exponent $\mathrm{\alpha = 4}$ for non-line-of-sight (NLoS) lossy environments. Without loss of generality, the small scale path fading model is scattered in a 2D circular symmetric Gaussian distribution which generalise the fading gain under exponential distribution with mean of $\mathrm{\lambda_{g} = 1}$ and the gateway received signal noise power is $\mathrm{\sigma_{n}^2 = -90}$ dB. The three pillars of the wireless network performance parameters are numbered as PPP user density $\mathrm{\lambda = [-60, -50]}$ dB, expected SINR quality limits in $\mathrm{\gamma^{th} = [-10, 0]}$ dB, and the range of network coverage probabilities $\mathrm{P_{cov} = [0, 1]}$. Now we analyse the variation of the firing power of IoT uplinks against the essential network metrics such as the connected user distance and three pillars. Then, the optimal power startgies were obtained from the proposed EBGT frameworks are validated with the expected network coverage performances through Monte-Carlo simulations utilizing realistic network dynamics. Furthermore, we make a performance comparison of the desired resource allocation protocol compared to fair practical baselines as follows. 

\subsection{Uplink Power Allocation against User Distance}
\begin{figure}[t]
\centering
\includegraphics[width=0.9\columnwidth, trim={2mm 2mm 9mm 6mm},clip]{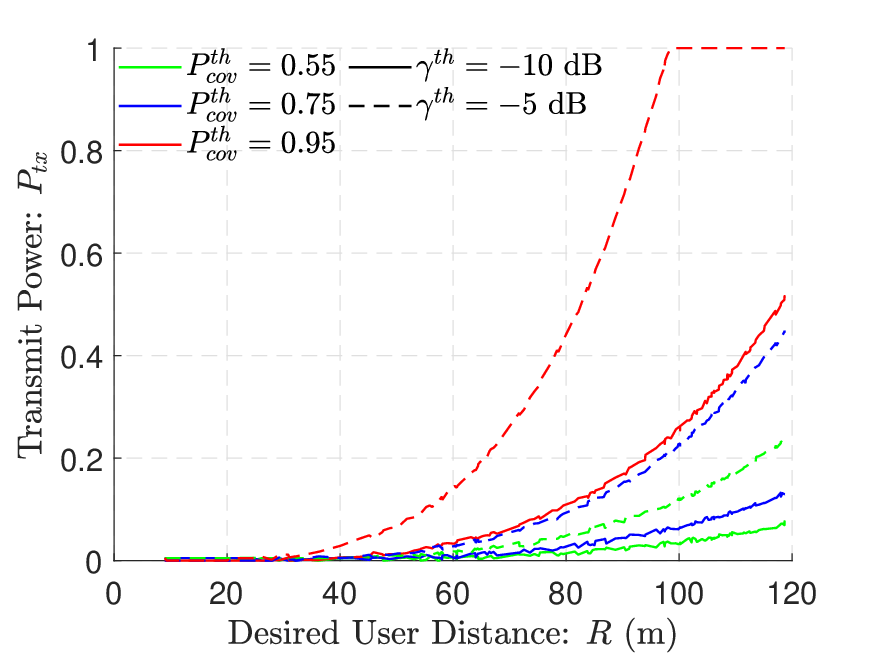}
\caption{Variation of normalized transmit power $\mathrm{\left(P_{tx}^{i} \text{ Vs } r_{i}\right)}$ of the author-desired IoT node $\mathrm{i \in \mathcal{N}}$, against the typical connected distance $\mathrm{r_{i}}$, for the set of network coverage bounds $\mathrm{P_{cov}^{th}} = \{0.55, 0.75, 0.95 \}$, and the set of SINR thresholds $\mathrm{\gamma^{th}} = \{-10, -5\}$ dB, under parameter settings: $\mathrm{\lambda_{g}=1}$, $\mathrm{\lambda=-60}$ dB, $\mathrm{\alpha=4}$, $\mathrm{\sigma_{n}^{2} = -90}$ dB, $\mathrm{\bar{R}=120}$~m, $\mathrm{Area=10}$~km$^2$.}
\label{Fig: Ptx_vs_ri}
\end{figure}
\figurename~\ref{Fig: Ptx_vs_ri} illustrates the variation of the normalized uplink firing power against the radiated distance from each author-desired IoT node to the connected gateway. The transmit power increases with the desired user distance to compensate for signal attenuation due to the power-law path loss. Therefore, the proposed EBGT framework addresses the fundamental power allocation task corresponding to the strength of the channel gain, apart from dealing with opponent node interferences. The power curves are uplifted diagonally for the tightened network requirements, such as demanding higher transmit power vectors to satisfy the greater network coverage expectations, and the SINR threshold bounds. For instance, IoT devices that are radiated with $\mathrm{r_{i} = 100}$ m and expecting SINR margin $\mathrm{\gamma^{th}=-10}$ dB, consume less than $\mathrm{10\% P_{tx}^{max}}$ transmit power portion for $\mathrm{55\%}$ successful uplinks however, demands approximately $\mathrm{15\% P_{tx}^{max}}$ to coverage $\mathrm{75\%}$ users in the IoT tier. In addition, IoT devices rely on a significant amount of transmit power levels to achieve higher channel throughput targets satisfying an advanced SINR margin $\mathrm{\gamma^{th} = -5}$ dB, represented by dashed lines. Quantitatively, the author-desired IoT nodes $\mathrm{i \in \mathcal{N}}$ that acquire a coverage probability of $\mathrm{95\%}$ with $\mathrm{\gamma^{th} = -10}$ dB \emph{(red-thick line)}, depending on a slightly similar power vector related to the IoT devices that expect $\mathrm{75\%}$ network coverage, though under a intensive channel throughput functioning on $\mathrm{\gamma^{th} = -5}$ dB \emph{(blue-dashed line)}. Furthermore, there is a substantial transmit power escalation for the IoT nodes expecting $\mathrm{95\%}$ network coverage and a strict SINR bound of $\mathrm{\gamma^{th} = -5}$ dB \emph{(red-dashed line)}, call for a saturated power level $\mathrm{P_{tx}^{max}}$ exceeding the connected distance of $\mathrm{r_{i} = 100}$ m.    
\subsection{Uplink Power Allocation against Three Pillars}
\begin{figure*}[t]
\centering
$\begin{array}{ccc}
\includegraphics[width=0.32\textwidth, trim={3mm 1mm 12mm 6mm},clip]{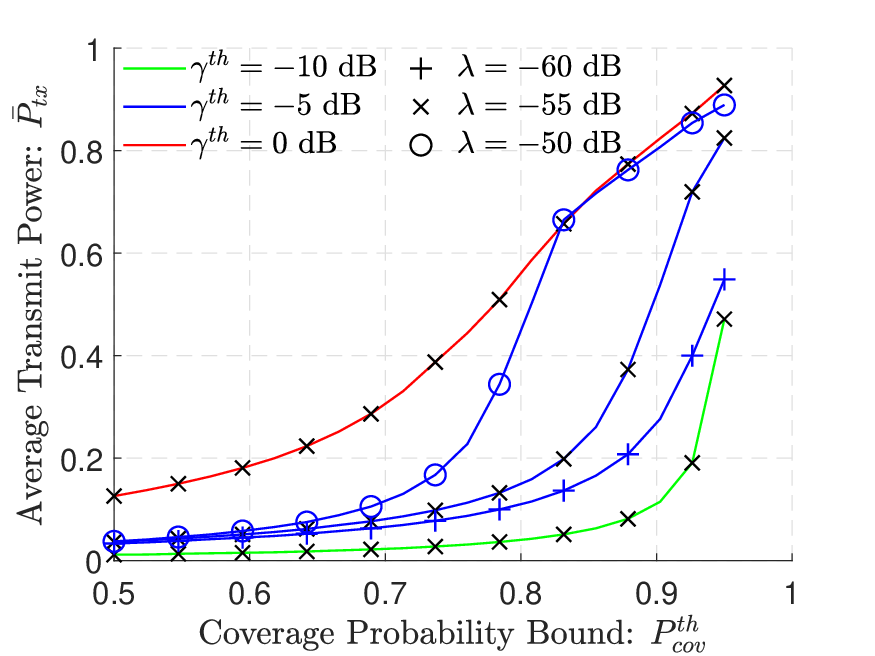} 
&
\includegraphics[width=0.32\textwidth, trim={3mm 1mm 12mm 6mm},clip]{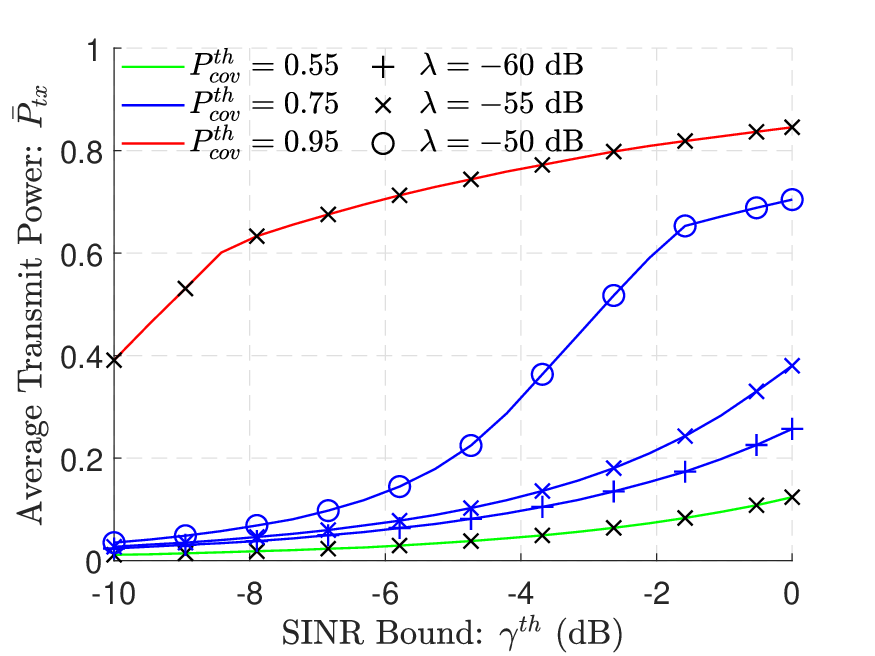}
&
\includegraphics[width=0.32\textwidth, trim={3mm 1mm 10mm 6mm},clip]{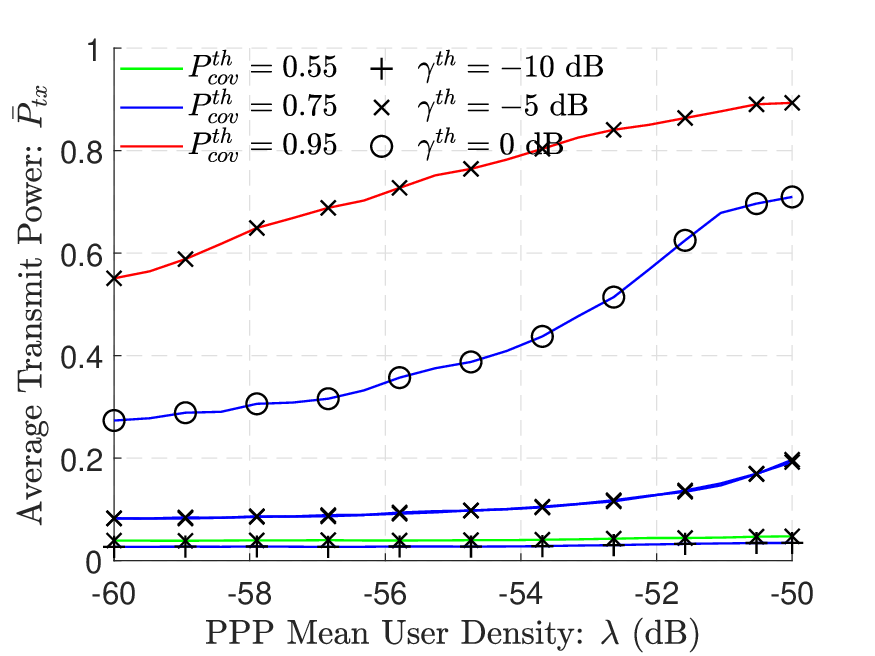} \\
\mbox{({\textit{a}}) $\mathrm{\bar{P}_{tx}}$ \text{ Vs } $\mathrm{P_{cov}^{th}}$} & \mbox{({\textit{b}}) $\mathrm{\bar{P}_{tx}}$ \text{ Vs } $\mathrm{\gamma^{th}}$} & \mbox{({\textit{c}}) $\mathrm{\bar{P}_{tx}}$ \text{ Vs } $\mathrm{\lambda}$}\\
\end{array}$
\caption{Variation of normalized average transmit power, (a) $\mathrm{\left(\bar{P}_{tx} \text{ Vs } P_{cov}^{th}\right)}$ against the set of network coverage probability bounds $\mathrm{P_{cov}^{th}}$, (b) $\mathrm{\left(\bar{P}_{tx} \text{ Vs } \gamma^{th}\right)}$ against the set of SINR thresholds $\gamma^{th}$, (c) $\mathrm{\left(\bar{P}_{tx} \text{ Vs } \lambda \right)}$, against the set of PPP mean user densities $\lambda$ of the IoT network, for the set of network coverage probability bounds $\mathrm{P_{cov}^{th}} = \{0.55, 0.75, 0.95 \}$ dB, the set of SINR thresholds $\mathrm{\gamma^{th}} = \{-10, -5, 0\}$ dB, and the set of PPP mean user densities $\mathrm{\lambda} = \{-60, -55, -50\}$ dB, under parameter settings: $\mathrm{\lambda_{g}=1}$, $\mathrm{\alpha=4}$, $\mathrm{\sigma_{n}^{2} = -90}$ dB, $\mathrm{\bar{R}=120}$~m, $\mathrm{Area=10}$~km$^2$.}
\label{Fig: Ptx_vs_Pcov_gamma_lambda}
\end{figure*}
Here, we discuss the average power decision profiles of the IoT-tiers versus three prominent pillars in the IoT environment, such as the expected network coverage, the target SINR threshold, and the average PPP node density. First, \figurename~\ref{Fig: Ptx_vs_Pcov_gamma_lambda}(a) represents the average firing power vectors $\mathrm{\bar{P}_{tx}}$ against the coverage probability thresholds $\mathrm{P_{cov}^{th}}$ for given parameter settings of SINR bounds and mean PPP user densities of the IoT network. In brief, IoT devices are paying a larger portion of the primary battery by increasing the number of successful transmissions to the connected gateway. Additionally, the rate of monotone power increment is elevated beyond the shoulder region of each plot, which differs for parameter settings. There are two major performance illustrations to summarize the parameter settings fairly, such as \emph{blue-lines with three different symbols} depicting mean transmit power actions for a fixed SINR bound $\mathrm{\gamma^{th}=-5}$ dB with increasing mean PPP user densities $\mathrm{\lambda}$ to rival the game, introducing a larger number of subscriptions to the network and targeting a unique channel throughput threshold. Secondly, uplink average power decisions against network coverage are presented via \emph{cross-symbol and three-color lines} for a fixed mean PPP user density $\mathrm{\lambda = -55}$ dB and SINR threshold parameter sub-settings $\mathrm{\gamma^{th}}$. This reveals the resource allocation performance mapping to heterogeneous data rate requirements in the IoT network for a static set of opponents. 

It is noticeable that the power increment plots are deviating at various rates towards the extreme conditions of parameter settings. For instance, transmit power growth rate drops beyond the network coverage targets, approximately $\mathrm{P_{cov}^{th} = 0.85}$ for the two IoT tiers that rely on parameter settings of $\mathrm{\{ \gamma^{th} = -5\text{ dB}, \lambda = -50 \text{ dB}\}}$ and $\mathrm{\{ \gamma^{th} = 0 \text{ dB}, \lambda = -55 \text{ dB}\}}$. This emphasis the ability of self-adaptiveness of the proposed EBGT resource allocation framework for the network dynamics, while reducing the firing strength to avoid unnecessary collisions to climb the network coverage ranks, though preserving higher channel throughput and interference challenges. Furthermore, these two IoT tiers share slightly similar power levels beyond the adaptive network coverage region, i.e., $\mathrm{P_{cov}^{th} = 0.85}$ for medium SINR margin with the desired highest mean PPP user density and vice versa. 

\figurename~\ref{Fig: Ptx_vs_Pcov_gamma_lambda}(b) shows the performance of transmit power allocation $\mathrm{\bar{P}_{tx}}$ with respect to second pillar: the SINR threshold margins $\mathrm{\gamma^{th}}$ for the parameter settings of coverage probabilities and mean PPP node densities. IoT devices demand higher transmission power values for greater SINR threshold bounds to preserve the reliability of the network. We outline the entire performance utilizing two sub-parameter settings, in particular \emph{blue-lines with three different symbols} portray power actions for diverse interference densities from $\mathrm{lambda = -60}$ dB to $\mathrm{lambda = -50}$ dB by $\mathrm{(\cross 10)}$ times enhancement of the network traffic for $\mathrm{P_{cov}^{th} = 0.75}$ coverage expectations. Subsequently, IoT players are boosting transmit power operations to enhance the number of successful transmissions for the intended $\mathrm{P_{cov}^{th}=0.55 \rightarrow0.95}$ alongside the fixed number of IoT opponent density $\mathrm{\lambda = -55}$ dB as demonstrated in \emph{cross-symbol and three color lines}. Moreover, IoT devices that secure massive network coverage $\mathrm{P_{cov}^{th} = 0.95, \lambda = -55}$ dB anticipate a substantial uplink strength than the intermediate coverage target of $\mathrm{P_{cov}^{th} = 0.75, \lambda = -50}$ dB, although there is a high volume of opponents. This reflects the intuitiveness of identifying player interactions in the IoT game model while making beliefs about opponents, which sustain critical wireless metrics. Simply, transmit power gains of IoT devices are monotonically increasing in parallel to higher SINR bounds $\mathrm{\gamma^{th}}$, though encouraged to lower the slope after a certain operating point, such as beyond $\mathrm{\gamma^{th} = -8.5}$ dB for \emph{cross-symbols on red-line} and about $\mathrm{\gamma^{th} = -1.5}$ dB for \emph{circular-symbols on blue-line} to mitigate the strength of the interference summation. Analogously, these two performance plots reveal that IoT players build the belief hierarchy, generating a substantial transmission power decrement when there are massive interference hands-on \emph{(higher mean PPP user densities, $\mathrm{\lambda}$)} and similarly applicable for another pair of \emph{cross-symbols on green line}: $\mathrm{P_{cov}^{th} = 0.55, \lambda = -50}$ dB with \emph{plus-symbols on blue line}: $\mathrm{P_{cov}^{th} = 0.75, \lambda = -60}$ dB. 

There is a mild growth for IoT power strategies against the third performance pillar specifically mean PPP user density $\mathrm{\lambda}$ as shown in \figurename~\ref{Fig: Ptx_vs_Pcov_gamma_lambda}(c) compared to previous figures. Following the same trend, we recap the entire range of the parameter settings across two primary ways such as \emph{blue-lines with three different symbols}, illustrating uplink power gains versus set of traffic densities for heterogeneous channel throughput margins $\mathrm{\gamma^{th} = -10 dB \rightarrow 0 dB}$ assited by $\mathrm{P_{cov}^{th} = 0.75}$ coverage probability. Concurrently, the transmit power controls versus mean opponent density for different network coverage goals from $\mathrm{P_{cov}^{th} = 0.55 \rightarrow 0.95}$ with SINR bound $\mathrm{\gamma^{th} = -5}$ dB are visualized through \emph{cross-symbol and three color lines}. There is not or slight power growth relative to IoT interference density $\mathrm{\lambda}$ for lower request of communication metrics such as $\mathrm{P_{cov}^{th} = 0.55, 0.75}$ and $\mathrm{\gamma^{th} = -10 dB, -5 dB}$. In contrast, IoT devices that satisfy $\mathrm{\gamma^{th} = -5}$ dB and forecasting $\mathrm{P_{cov}^{th} = 0.95}$ back-to-back transmissions \emph{(cross-symbols on red line)} escalate power requirements substantially more than the less network coverage $\mathrm{P_{cov}^{th} = 0.75}$ expectations though preserving tighten throughput bounds $\mathrm{\gamma^{th} = 0}$ dB. This further reflects that IoT players mitigate the growth rate of transmit power actions to weaken the interference addition in SINR bounds $\mathrm{\gamma^{th}}$ alongside dealing with massive number of collisions $\mathrm{\lambda \rightarrow -50}$ dB.

In conclusion, the proposed EBGT resource allocation approach regulates the transmit power operating point fairly corresponding to three prominent desired pillars, such as more encouragement to enhance the power growth rate by forecasting greater coverage probabilities $\mathrm{P_{cov}^{th}}$. However, IoT players tailor the belief hierarchy about competitive interference opponents while degrading the power operating point and are reluctant to escalate the transmit power, facilitating critical IoT users who are more prone for extreme number of collisions $\mathrm{\lambda}$ and greater SINR bounds $\mathrm{\gamma_{th}}$.      

\subsection{Evaluation of Network Coverage}
\begin{figure*}[t]
\centering
$\begin{array}{ccc}
\includegraphics[width=0.45\textwidth, trim={0mm 0mm 0mm 0mm},clip]{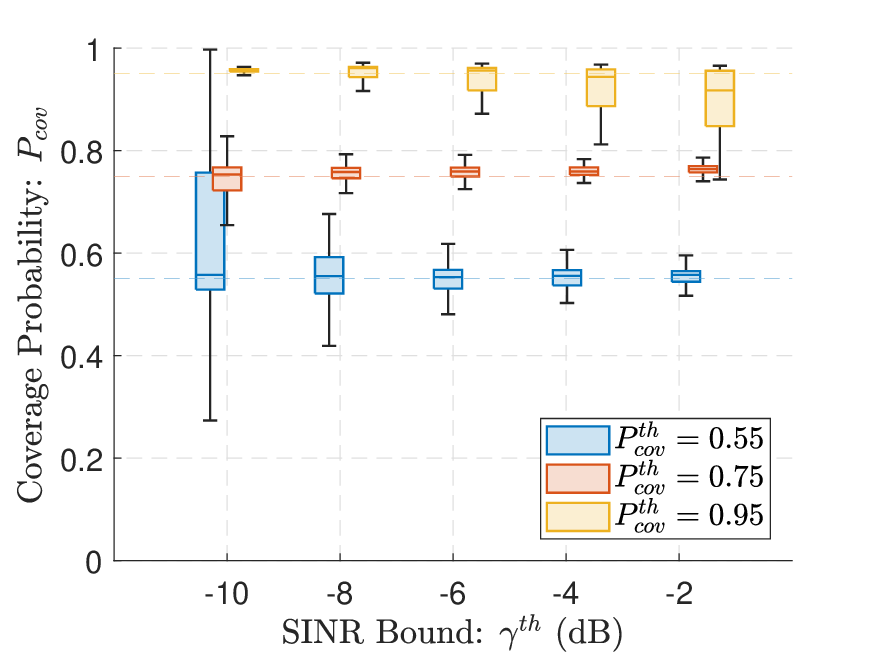} 
&
\includegraphics[width=0.45\textwidth, trim={0mm 0mm 0mm 0mm},clip]{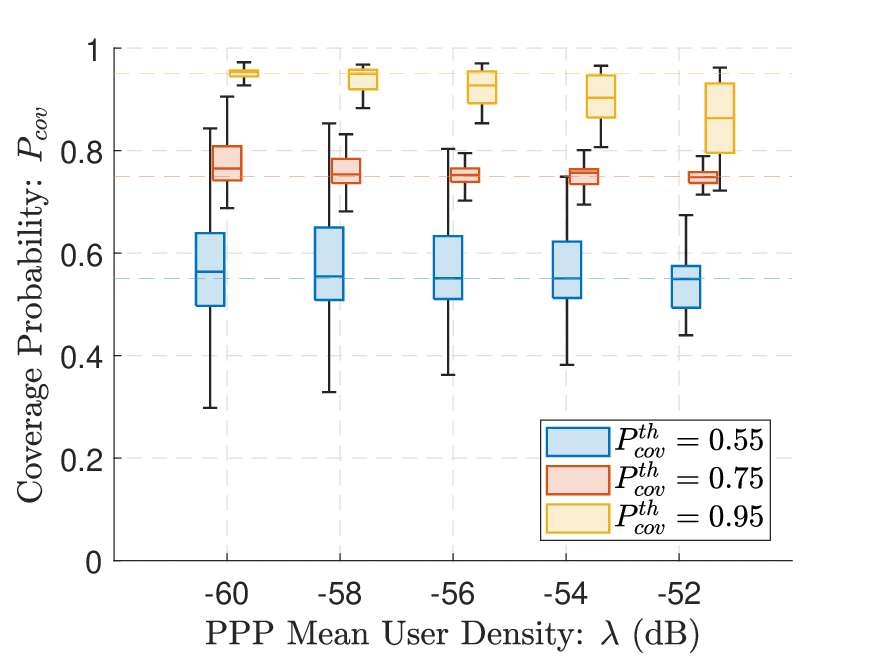} \\
\mbox{({\textit{a}}) $\mathrm{P_{cov}}$ \text{ Vs } $\mathrm{\gamma^{th}}$, $\mathrm{\lambda = -55}$ dB} & \mbox{({\textit{b}}) $\mathrm{P_{cov}}$ \text{ Vs } $\mathrm{\lambda}$, $\mathrm{\gamma^{th} = -5}$ dB}\\
\end{array}$
\caption{Evaluation of coverage probability, (a) $\mathrm{\left(P_{cov} \text{ Vs } \gamma^{th}\right)}$, against the set of SINR thresholds $\gamma^{th}$, (b) $\mathrm{\left(P_{cov} \text{ Vs } \lambda \right)}$, against the set of PPP mean user densities $\lambda$ of the IoT network, for the set of network coverage probability bounds $\mathrm{P_{cov}^{th}} = \{0.55, 0.75, 0.95 \}$ dB, under parameter settings: $\mathrm{\lambda_{g}=1}$, $\mathrm{\alpha=4}$, $\mathrm{\sigma_{n}^{2} = -90}$ dB, $\mathrm{\bar{R}=120}$~m, $\mathrm{Area=10}$~km$^2$.}
\label{Fig: Pcov_vs_gamma_lambda}
\end{figure*}
Thereafter, firing optimal transmit-power actions via the proposed EBGT resource allocation framework, the coverage probability of the entire IoT network is evaluated through Monte-Carlo simulation, accounting for realistic network constraints. We validate the expected coverage performance using a candle plot diagram, where the bottom and top edges of the filled body explain the 1st and 3rd quartiles of the coverage distribution, respectively. The median represents the desired coverage target achieved by IoT devices more frequently then, the lowest and highest outliers are indicated by wicks of candles. 

\figurename~\ref{Fig: Pcov_vs_gamma_lambda}(a) depicts the coverage performance of the IoT network densified by $\mathrm{\lambda = -55}$ dB compared to SINR thresholds for three different coverage expectations. The lower-expected network coverage $\mathrm{P_{cov}^{th} = 0.55}$ bound \emph{(blue candles)} contains a higher degree of freedom (DoF) to fluctuate in a broader coverage region for relaxed SINR $\mathrm{\gamma^{th} \rightarrow -10}$ dB margins. However, most of the IoT devices could exceed the expected network coverage, which shows the median is aligned to $\mathrm{P_{cov}^{th} = 0.55}$, surpassing the candle body beyond the bound. On the other hand, IoT players regulate the transmit power operating point according to the beliefs about the opponent interfering nodes and strict SINR bounds $\mathrm{\gamma^{th} \rightarrow 0}$ dB. Hence, the DoF of the network coverage distribution is squeezed toward the desired  $\mathrm{P_{cov}^{th}}$ against the higher channel throughput values. There is an aggregate performance for successful transmissions in the network under $\mathrm{P_{cov}^{th} = 0.75}$, \emph{(red candles)}, and preserve the given target coverage. In addition, IoT gateways observe received signals successfully for higher coverage probabilities $\mathrm{P_{cov}^{th} = 0.95}$, though loose SINR bounds \emph{(amber candles)}. Nevertheless, some of the IoT uplinks' performance is slightly degraded to maintain the desired margins for challenging significant SINR targets on top of the higher coverage probabilities. At this stage, the proposed EBGT protocol with inter-epistemic strategies is susceptible to refluence the slope of the transmit power strategies as shown in \emph{(red line with cross symbols)} of \figurename~\ref{Fig: Ptx_vs_Pcov_gamma_lambda}(b) and mitigate the interference summation.

The coverage probability metric versus interference traffic in the IoT network for channel throughput requirement of $\mathrm{\gamma^{th} \rightarrow -5}$ dB is discussed in \figurename~\ref{Fig: Pcov_vs_gamma_lambda}(b). There is a considerable DoF of coverage probability that varies in a broader region \emph{(blue candles)} for IoT devices targeting $\mathrm{P_{cov}^{th} = 0.55}$, yet manage to satisfy the threshold for the vast majority. Moreover, IoT players prioritize the intra-epsitemic belief layers while enhancing the rate of the transmit power curve, \emph{(blue line with cross symbols)} of \figurename~\ref{Fig: Ptx_vs_Pcov_gamma_lambda}(c) together with the increasing number of opponent nodes $\mathrm{\lambda \rightarrow -50}$ dB in the network. The associated achievable coverage probabilities $\mathrm{P_{cov}^{th} = 0.75}$ are presented using \emph{(red candles)} that sustain and compress around the coverage bound, gaining higher power growth in parallel to the massive number of interference nodes in the network. Although the IoT players behave well in superior coverage $\mathrm{P_{cov}^{th} = 0.95}$, dealing with a smaller number of neighbor nodes \emph{(amber candles)},  certain successful deliveries deviate slightly from the expected coverage for rigorous network traffic $\mathrm{\lambda \rightarrow -50}$ dB. However, the proposed EBGT model inherently calibrates the inter-epistemic belief layers, whereas reducing the power growing slope \emph{(red line with cross symbols)} after nearly $\mathrm{\lambda \rightarrow -58}$ dB in \figurename~\ref{Fig: Ptx_vs_Pcov_gamma_lambda}(c). Therefore, uplinks from the most desired IoT devices could reach the tagged gateway appropriately, avoiding the strength of the collision against frequent interferers in the network.       
\subsection{Performance Comparison Metrics}
We compare the resource allocation capability of the proposed beliefs-inspired EBGT framework with two benchmarks, called fixed power control (FPC) and stochastic non-cooperative power control (SNCPC). In the FPC approach, IoT devices select a static power level that is the maximum of the proposed EBGT framework $\mathrm{\textbf{P}_{rx}^{FPC} = Max_{\forall i \in \mathcal{N}} \textbf{P}_{tx}^{EBGT}}$ against the SINR bound $\mathrm{\gamma^{th}}$ to energize the uplink signals. And, SNCPC considers an independent PPP IoT user distribution and selects the best transmit power decision to satisfy the given SINR margin in a non-cooperative way. To ensure a fair comparison, benchmark schemes are selected to follow a similar algorithmic paradigm as the proposed approach, thereby avoiding biases due to heterogeneous modeling assumptions. In particular, artificial neural network (ANN)-based methods are sensitive to the size and randomness of the dataset, with training time, classical methods rely on deterministic data assumptions, and distributed feedback-based approaches introduce additional idle overhead due to signaling and coordination.

\begin{figure*}[t]
\centering
$\begin{array}{ccc}
\includegraphics[width=0.32\textwidth, trim={3mm 1mm 7mm 6mm},clip]{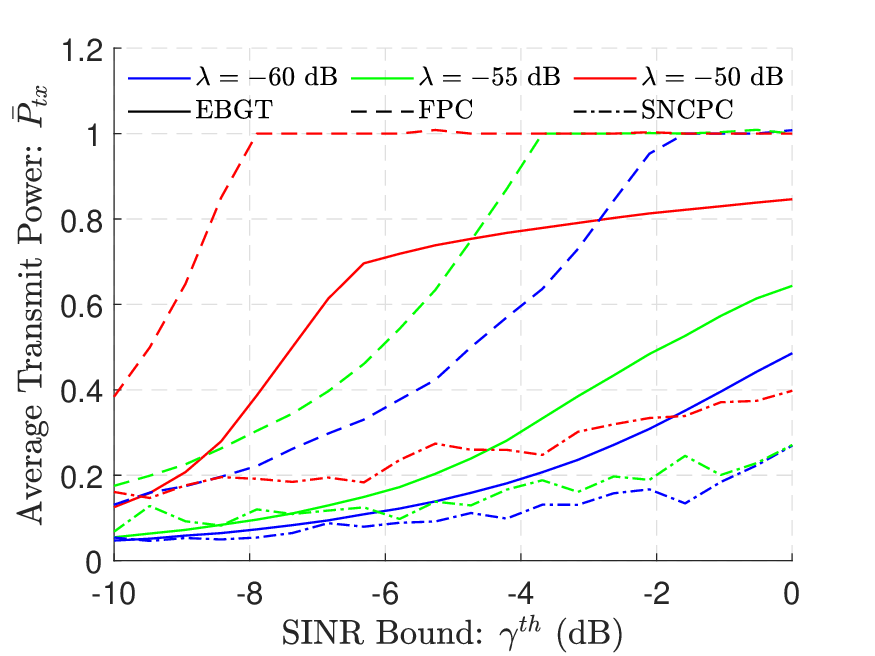} 
&
\includegraphics[width=0.32\textwidth, trim={3mm 1mm 7mm 6mm},clip]{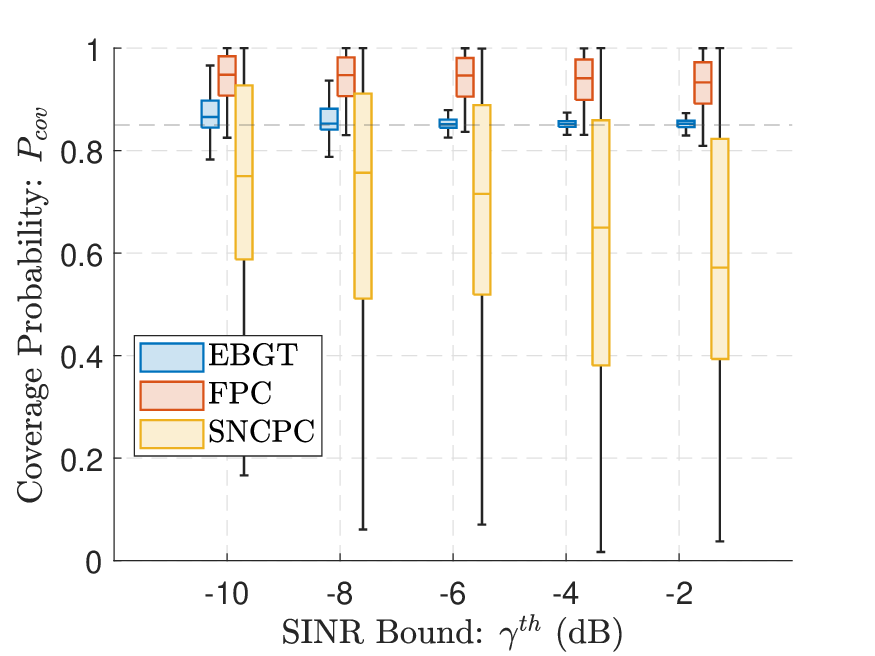}
&
\includegraphics[width=0.32\textwidth, trim={3mm 1mm 7mm 6mm},clip]{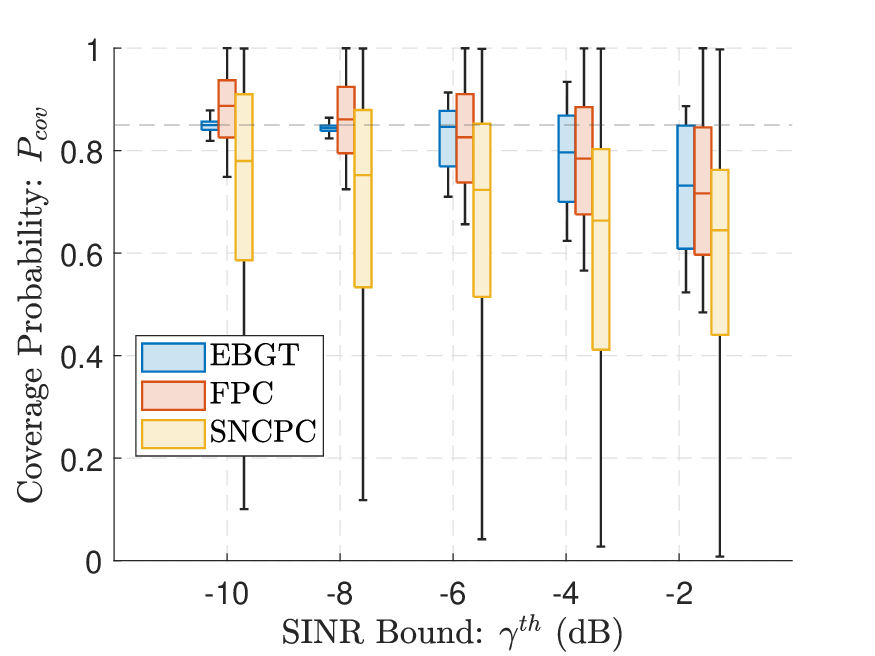} \\
\mbox{({\textit{a}}) $\mathrm{\bar{P}_{tx} \text{ Vs } \gamma^{th}}$} & \mbox{({\textit{b}}) $\mathrm{P_{cov} \text{ Vs } \gamma^{th}, \lambda = -60}$ dB} & \mbox{({\textit{c}}) $\mathrm{P_{cov} \text{ Vs } \gamma^{th}, \lambda = -50}$ dB}\\
\end{array}$ 
\caption{(a) Variation of normalized average transmit power $\mathrm{\left(\bar{P}_{tx} \text{ Vs } \gamma^{th}\right)}$ \& (b), (c) Evaluation of network coverage probability $\mathrm{\left(P_{cov} \text{ Vs } \gamma^{th}\right)}$, against the set of SINR threshold bounds $\mathrm{\gamma^{th}}$ for the specified network coverage bound $\mathrm{P_{cov}^{th}} = 0.85$, under parameter settings: $\mathrm{\lambda_{g}=1}$, $\mathrm{\alpha=4}$, $\mathrm{\sigma_{n}^{2} = -90}$ dB, $\mathrm{\bar{R}=120}$~m, $\mathrm{Area=10}$ km$^2$.}
\label{Fig: Benchmarks}
\end{figure*}
\figurename~\ref{Fig: Benchmarks}(a) compares the average transmission power vectors of the IoT network versus the channel throughput requirement $\mathrm{\gamma_{th}}$ of the proposed EBGT approach against the benchmarks of FPS and SNCPC. There are three distinct PPP mean user distributions $\mathrm{\lambda = \{-60, -55, -50\}}$ dB and an increased number of possible collisions respectively at the IoT gateway. Although demanding lesser power amount, SNCPC baseline would not be able to capture IoT node interactions by two means, such as \emph{working in the high-power region}: lack of controllability to ceil the power operating point to reduce the interference summation that would benefit neighbor uplinks, and \emph{working in the low-power region}: inactivity to floor the power operating point results in fragile uplinks at the gateway oppose to neighbour firings. Likewise, FPC locks the uplink strength mapping to the required SINR target and utilizes an extensive amount of battery percentage for each IoT device. In parallel, the uncontrolled power growing slope tends towards rapid saturation at $\mathrm{P_{tx}^{max}}$, indicated nearly $\mathrm{\gamma^{th} = -1.5}$ dB for $\mathrm{\lambda = -60}$ dB, $\mathrm{\gamma^{th} = -3.5}$ dB for $\mathrm{\lambda = -55}$ dB, $\mathrm{\gamma^{th} = -8}$ dB for $\mathrm{\lambda = -50}$ dB. The introduced EBGT adopts a balanced uplink power strategy and adapts to decrease the power growing rate smoothly for extreme network metrics toward $\mathrm{\gamma^{th} \rightarrow 0}$ dB for $\mathrm{\lambda \rightarrow -50}$ dB.       

Nevertheless, the SNCPC approach is comfortable for lower transmit power gains, there is a broader deviation for expected coverage probability, illustrated by \emph{amber candles} in \figurename~\ref{Fig: Benchmarks}(b) and (c) for two different mean PPP user densities $\mathrm{\lambda = -60}$ dB and $\mathrm{\lambda = -60}$ dB. A few IoT nodes that sustain higher transmit power with strong channel gains achieves tremendous network coverage $\mathrm{P_{cov}^{i} \rightarrow 1}$, however, this may penalize adjacent IoT uplinks $\mathrm{P_{cov}^{j \in \mathcal{N \setminus i}}}$ experiencing weaker channel gains and lower transmission strength, and vice versa. The generalized EBGT protocols \emph{blue candles} guarantee the desired coverage probability $\mathrm{P_{cov}^{tx} = 0.85}$, and FPC acquire better coverage results \emph{red candles} for $\mathrm{\lambda = -60}$ dB. Conversely, FPC demands excessive power levels $\mathrm{70\%P_{tx}^{max}}$ for the uplink than the proposed EBGT approach at $\mathrm{\gamma^{th} = -2}$ dB, \emph{(massive gap between blue thick and dashed lines)}. Additionally, coverage performance of FPC is less resilient for intensive network constraints $\mathrm{(\gamma^{th} \rightarrow -5 \text{ dB and } \lambda = -50 \text{ dB})}$, and is more energy hunger for the uplink nearly $\mathrm{20\% P_{tx}^{max}}$ than the proposed EBGT, at $\mathrm{\gamma^{th} = -2}$ dB, \emph{(massive gap between red thick and dashed lines)}. Optimistically, the EBGT resource allocation framework is well-adapted to notify extreme parameter limitations and take a turn nearly $\mathrm{\gamma^{th} = -6.5}$ dB to suppress the power growing rate for $\mathrm{\lambda = -50}$ dB, \emph{(red thick line)}. This would diminish the interference strength of the network to combat the surge in traffic. Hence, \figurename~\ref{Fig: Benchmarks} reflects the importance of the belief-based rational resource allocation protocols that could secure the achievable network coverage accepting minimal transmit power magnitudes.

Furthermore, virtaul cooperation among agents follwoing inter and intra-epistemic belief layers would converge toward the global equilibrium employing less number of calculation generalized lightweight algorithm and compatible with low SWaP IoT processors. 

\section{Conclusion}
We proposed an epistemology-aided Bayesian game theory (EBGT) resource allocation protocol for stochastically distributed IoT networks in the presence of neighbor interference. Each author-desired IoT device has a discipline of IoT uplink power regulation against the opponent user firings, while satisfying the coverage performance of the entire network. We consider realistic user dynamics through a PPP distributed stochastic IoT network, and challenges of unaware CSI among entities. To address the critical power control problem capturing IoT node behaviors, we employed Bayesian game theory as the kernel of the framework. Here, virtual cooperation among agents across inter- and intra-epistemic belief layers would converge toward the global equilibrium using a smaller number of calculations with a generalized lightweight algorithm, and be compatible with low SWaP IoT processors. Simulations show competitive uplink management and strong reliability, while enhancing the power efficiency against channel throughput bounds. The proposed EBGT architecture establishes a computationally lean and adaptive framework for scalable deployment in dense IoT networks. By enabling optimal power allocation strategies, it effectively mitigates cross-channel collisions while maintaining high network reliability and resource efficiency. Future work will analyze the specific utility distribution for higher-order statistical (HoS) derivations while extracting unique interactive features among users to validate the EBGT approach on empirical network metrics in CPS.

\section{Appendix} \label{Sec: appendix}
\subsection{Coverage Probability} \label{appendix: coverage_prob}
\begin{align} \label{Eq: coverage_prob_integral}
    \begin{split}
        \mathrm{P_{cov}^{i}} &= \mathrm{ \int_{0}^{\bar{R}} \Pr \left( \gamma_{i} \geq \gamma^{th} \big| r_{i} \right) . f_{R}(r_{i}|\bar{R}) dr_{i}}
        \\
        & \text{Using Rayleigh distance distribution from \eqref{Eq: Rayl_PDF_PPP_truncated},}
        \\
        &= \mathrm{\bigintssss_{0}^{\bar{R}} \Pr \left( \frac{{g}_i {p}_i  r_{i}^{-\alpha}}{\mathcal{I}+\sigma_{n,i}^2} \geq \gamma^{th} \Bigg| r_{i} \right) . \left( \frac{2 \pi \lambda r_{i} e^{- \pi \lambda r_{i}^2}}{1 - e^{-\pi \lambda \bar{R}^2}} \right)  dr_{i}}
        \\
        &= \mathrm{ \frac{2 \pi \lambda}{1 - e^{-\pi \lambda \bar{R}^2}} \hspace{-2mm}\bigintssss_{0}^{\bar{R}} \hspace{-2mm} e^{- \pi \lambda r_{i}^2}. \underbrace{\Pr \left( \frac{{g}_i {p}_i  r_{i}^{-\alpha}}{\mathcal{I}+\sigma_{n,i}^2} \geq \gamma^{th} \Bigg| r_{i} \right)}_{\Pr \left( \cdot\right)} r_{i} dr_{i}}
    \end{split}
\end{align}
The inner probability term with respect to $\mathrm{r_{i}}$ and $\mathrm{\mathcal{I}}$
\begin{align} \label{Eq: inner_prob_expectation}
    \begin{split}
        \mathrm{\Pr \left( \cdot\right)} &= \mathrm{\mathbb{E}_{\mathcal{I}} \left[ \Pr \left( {g}_i \geq \gamma^{th} {p}_i^{-1}  r_{i}^{\alpha} \left( \mathcal{I}+\sigma_{n,i}^2 \right) \big| r_{i}, \mathcal{I} \right) \right]}
        \\
        & \text{Small scale path fading, $\mathrm{G_{i} \sim exp(\lambda_g)}$}
        \\
        &= \mathrm{\mathbb{E}_{\mathcal{I}} \left[ e^{-\lambda_g \gamma^{th} {p}_i^{-1}  r_{i}^{\alpha} \left( \mathcal{I}+\sigma_{n,i}^2 \right)} \Big| r_{i}, \mathcal{I}  \right]}
        \\
        &= \mathrm{e^{- \lambda_g \gamma^{th} {p}_i^{-1}  r_{i}^{\alpha} \sigma_{n,i}^2 }. \mathcal{L}_{\mathcal{I}} \left( \lambda_g \gamma^{th} {p}_i^{-1}  r_{i}^{\alpha} \right) }
    \end{split}
\end{align}

\subsection{Laplace interference characterization, $\mathrm{\mathcal{L}_{\mathcal{I}}} = \mathrm{\mathbb{E}_{\mathcal{\mathcal{I}}}\left[ e^{-\zeta \mathcal{I}} \right]}$} \label{appendix: Laplace_interfernce_characetrization}
\begin{align} \label{Eq: intf_charac}
    \begin{split}
        \mathrm{\mathcal{L}_{\mathcal{I}}} &= \mathrm{\mathbb{E}_{\mathcal{I}}\left[ \exp{\left( -\zeta \sum_{j \in \mathcal{N} \setminus i} \sum_{\mathbf{x} \in \mathbf{\Phi}_{\mathcal{I}_j}} p_{j} g_{\mathbf{x}} \left\| \mathbf{x} - \mathbf{b}_{i} \right\|^{-\alpha} \right)} \right]}
        \\
        &= \mathrm{\mathbb{E}_{G_{\mathbf{x}}, \mathbf{X}}\left[ \prod_{j \in \mathcal{N} \setminus i} \prod_{\mathbf{x} \in \mathbf{\Phi}_{\mathcal{I}_j}} \exp{\left( -\zeta  p_{j} g_{\mathbf{x}} \left\| \mathbf{x} - \mathbf{b}_{i} \right\|^{-\alpha} \right)} \right]}
        \\
        & \text{Consider the independence of $\mathrm{G_{\mathbf{x}}}$ across $\mathrm{\mathbf{\Phi}_{\mathcal{I}_j}}$,}
        \\
        &= \mathrm{ \prod_{j \in \mathcal{N} \setminus i} \mathbb{E}_{\mathbf{\Phi}_{\mathcal{I}_j}}\left[ \prod_{\mathbf{x} \in \mathbf{\Phi}_{\mathcal{I}_j}} \mathbb{E}_{G_{\mathbf{x}}} \left\{ \exp{\left( -\zeta  p_{j} g_{\mathbf{x}} \left\| \mathbf{x} - \mathbf{b}_{i} \right\|^{-\alpha} \right)} \right\} \right]}
        \\
         & \text{Since $\mathrm{G_{\mathbf{x}} \sim exp(\lambda_g)}$, using the moment generating function,}
         \\
         &= \mathrm{\prod_{j \in \mathcal{N} \setminus i} \mathbb{E}_{\mathbf{\Phi}_{\mathcal{I}_j}}\left[ \prod_{\mathbf{x} \in \mathbf{\Phi}_{\mathcal{I}_j}} \frac{\lambda_g}{\lambda_g + \zeta p_{j} \left\| \mathbf{x} - \mathbf{b}_{i} \right\|^{-\alpha} } \right]}
         \\
         & \text{Apply probability generating functional (PGFL), \cite{Primer_Dhillon}}\\
         & \text{over 2D PPP tiers,}
         \\
         &= \mathrm{\prod_{j \in \mathcal{N} \setminus i} \exp{ \left[ -\lambda_{j} \bigintssss_{\mathbb{R}^2} \left( 1 - \frac{\lambda_g}{\lambda_g + \zeta p_{j} \left\| \mathbf{x} - \mathbf{b}_{i} \right\|^{-\alpha} } \right) d\mathbf{x} \right] } }
         \\
         & \text{Simplify from Cartesian to Polar coordinates,}
         \\
         &= \mathrm{\exp{ \left[ -2 \pi \displaystyle \sum_{j \in \mathcal{N} \setminus i} \lambda_{j} \underbrace{\mathrm{\bigintssss_{0}^{\infty} \left(\frac{\zeta p_{j} r^{-\alpha}}{\lambda_g + \zeta p_{j} r^{-\alpha} } \right) r dr}}_{J} \right] }}
         \\
         & \text{Using substitution of $\mathrm{t = \zeta p_{j} r^{-\alpha}/\lambda_g}$ for the integral,}
         \\
         & \text{then from the properties of Beta function, \cite{imprpoer_integral}}
         \\
         \mathrm{J} &= \mathrm{\frac{\left(\zeta p_{j}/\lambda_g\right)^{2/\alpha}}{\alpha} \underbrace{\bigintssss_{0}^{\infty} \left(\frac{t^{-2/\alpha}}{1 + t } \right) dt}_{\underbrace{\Gamma\left(1-2/\alpha \right)\Gamma\left(2/\alpha \right)}_{\times\,(2/\alpha)\,=\,\Upsilon(\alpha)}} }
         \\
         \mathrm{\mathcal{L}_{\mathcal{I}}} &= \mathrm{\exp\left( -\pi \lambda (s/\lambda_g)^{2/\alpha} \Upsilon(\alpha) \mathbb{M}_{\frac{2}{\alpha}} \left[ p_{j} \right] \right)}
         \\
         & \text{where $\mathrm{\lambda_{j} = \lambda / (N-1)}$ and $\mathrm{\mathbb{M}_{\frac{2}{\alpha}} \left[ p_{j} \right] = \frac{1}{N-1} \sum_{j \in \mathcal{N} \setminus i}  p_{j}^{\frac{2}{\alpha}} }$}
         \\
         & \text{with } \mathrm{s=\zeta.}
    \end{split}
\end{align}

\subsection{Utility Concavity} \label{appendix: cov_prob_concav_interaction}
The derived coverage probability \eqref{Eq: covergae_probability_derived} is rewritten, 
\begin{align} \label{Eq: cov_prob_rewritten}
    \begin{split}
    \mathrm{P_{cov}^{i}} &= \mathrm{C_{1} \bigintssss_{0}^{\bar{R}} r_{i} e^{- \Theta(p_{i})} dr_{i}}  \text{, where}
    \\
    \mathrm{C_{1} = \frac{2 \pi \lambda}{1 - e^{-\pi \lambda \bar{R}^2}}} &;
    \mathrm{\Theta(p_{i})} = \mathrm{\pi \lambda \left( 1 + C_{2} {p}_i^{-2/\alpha}  \right) r_{i}^2 + C_{3} {p}_i^{-1}  r_{i}^{\alpha}}
    \\
    \mathrm{C_{2}} &= \mathrm{\left(\gamma^{th} \right)^{2/\alpha} \Upsilon(\alpha) \mathbb{M}_{\frac{2}{\alpha}} \left[ p_{j} \right]};
    \mathrm{C_{3}} = \mathrm{\lambda_g \gamma^{th} \sigma_{n,i}^2}
    \end{split}
\end{align}
\begin{align} \label{Eq: cov_prob_first_derivative}
    \begin{split}
    \mathrm{\left(P_{cov}^{i}\right)_{p_{i}} = \frac{\partial P_{cov}^{i}}{\partial p_{i}}} = \mathrm{-C_{1} \bigintssss_{0}^{\bar{R}} r_{i} e^{- \Theta} (\Theta)_{p_{i}} dr_{i}}
    \\
    \mathrm{(\Theta)_{p_{i}} = \frac{\partial \Theta(p_{i})}{\partial p_{i}}} = \mathrm{-\frac{2 \pi \lambda C_{2}} {\alpha} p_{i}^{-\frac{2}{\alpha}-1} r_{i}^{2} - C_{3} {p}_i^{-2}  r_{i}^{\alpha}}
    \end{split}
\end{align}
$\therefore \mathrm{\left(P_{cov}^{i}\right)_{p_{i}} > 0 \Rightarrow P_{cov}^{i}}$ is monotonically increasing.
\begin{align} \label{Eq: cov_prob_second_derivative}
    \begin{split}
    \mathrm{\left(P_{cov}^{i}\right)_{p_{i},p_{i}} = \frac{\partial^{2} P_{cov}^{i}}{\partial p_{i}^{2}} = C_{1} \bigintssss_{0}^{\bar{R}} r_{i} e^{- \Theta} \left[ (\Theta)_{p_{i}}^{2} - (\Theta)_{p_{i}p_{i}} \right] dr_{i}}
    \\
    \mathrm{(\Theta)_{p_{i}p_{i}} = \mathrm{\frac{2 \pi \lambda C_{2}} {\alpha} \left(  1+\frac{2}{\alpha} \right)p_{i}^{-\frac{2}{\alpha}-2} r_{i}^{2} + 2C_{3} {p}_i^{-3}  r_{i}^{\alpha}}}
    \\
    \therefore \mathrm{\left(P_{cov}^{i}\right)_{p_{i}p_{i}} < 0; \quad (\Theta)_{p_{i}}^{2} \ll (\Theta)_{p_{i}p_{i}}, \quad \left(\lambda, \sigma_{n,i}^2 \ll \right)}
    \\
    \mathrm{P_{cov}^{i}} \text{ is strictly concave.}
    \end{split}
\end{align}

\bibliographystyle{IEEEtran}
\bibliography{IEEEabrv,ref}
\end{document}